\documentclass[sigconf]{acmart}

\usepackage{amsmath}
\usepackage{amsfonts}
\usepackage{graphicx}
\usepackage{xcolor}
\usepackage{subfigure}
\usepackage{enumitem}
\usepackage{multirow}
\usepackage[tight]{units}

\AtBeginDocument{%
  \providecommand\BibTeX{{%
    \normalfont B\kern-0.5em{\scshape i\kern-0.25em b}\kern-0.8em\TeX}}}

\setcopyright{acmcopyright}
\copyrightyear{2021}
\acmYear{2021}
\setcopyright{acmcopyright}\acmConference[MICRO '21]{MICRO-54: 54th Annual IEEE/ACM International Symposium on Microarchitecture}{October 18--22, 2021}{Virtual Event, Greece}
\acmBooktitle{MICRO-54: 54th Annual IEEE/ACM International Symposium on Microarchitecture (MICRO '21), October 18--22, 2021, Virtual Event, Greece}
\acmPrice{15.00}
\acmDOI{10.1145/3466752.3480041}
\acmISBN{978-1-4503-8557-2/21/10}

\begin{document}

\title{SMART: A Heterogeneous Scratchpad Memory Architecture for Superconductor SFQ-based Systolic CNN Accelerators}

\author{Farzaneh Zokaee}
\email{fzokaee@iu.edu}
\affiliation{%
\institution{Indiana University}
\city{Bloomington}
\country{USA}
}

\author{Lei Jiang}
\email{jiang60@iu.edu}
\affiliation{%
\institution{Indiana University}
\city{Bloomington}
\country{USA}
}

\renewcommand{\shortauthors}{Zokaee and Jiang, et al.}

\begin{abstract}
Ultra-fast \& low-power superconductor single-flux-quantum (SFQ)-based CNN systolic accelerators are built to enhance the CNN inference throughput. However, shift-register (SHIFT)-based scratchpad memory (SPM) arrays prevent a SFQ CNN accelerator from exceeding 40\% of its peak throughput, due to the lack of random access capability. This paper first documents our study of a variety of cryogenic memory technologies, including Vortex Transition Memory (VTM), Josephson-CMOS SRAM, MRAM, and Superconducting Nanowire Memory, during which we found that none of the aforementioned technologies made a SFQ CNN accelerator achieve high throughput, small area, and low power simultaneously. Second, we present a heterogeneous SPM architecture, SMART, composed of SHIFT arrays and a random access array to improve the inference throughput of a SFQ CNN systolic accelerator. Third, we propose a fast, low-power and dense pipelined random access CMOS-SFQ array by building SFQ passive-transmission-line-based H-Trees that connect CMOS sub-banks. Finally, we create an ILP-based compiler to deploy CNN models on SMART. Experimental results show that, with the same chip area overhead, compared to the latest SHIFT-based SFQ CNN accelerator, SMART improves the inference throughput by $3.9\times$ ($2.2\times$), and reduces the inference energy by $86\%$ ($71\%$) when inferring a single image (a batch of images).
\end{abstract}



\maketitle

\section{Introduction}
\label{s:intro}

Deep learning has been the dominant approach to solving a wide variety of problems such as computer vision~\cite{Krizhevsky:NIPS2012}, natural language processing, and recommender systems. However, an inference of convolutional neural networks (CNNs) requires a multitude of com-puting-intensive convolutions. For instance, an AlexNet inference \cite{Krizhevsky:NIPS2012} costs 1.5 billion multiply-accumulate (MAC) operations involving 61 million parameters. As the era of Moore's law draws to a close, recent work~\cite{Ishida:MICRO2020} builds a systolic CNN accelerator, SuperNPU, to process CNN inferences by superconductor SFQ logic. The SFQ technology~\cite{Mukhanov:TAS2011,Yorozu:PCS2002} enables a low-level voltage impulse-driven switching, so that SFQ-based designs can achieve extremely high frequency (e.g., $\sim\unit[70]{GHz}$) but consume only tiny energy (e.g., $\unit[10^{-19}]{J}$ per switching). SuperNPU~\cite{Ishida:MICRO2020} is designed to run at $\unit[52]{GHz}$ by consuming only $\unit[1.9]{W}$ power. Compared to the state-of-the-art (SOTA) CMOS TPU~\cite{Jouppi:ISCA2017}, SuperNPU improves the batch inference throughput of various CNNs by $23\times$.

\textit{Unfortunately, the inference throughput of SFQ-based systolic CNN accelerators is seriously limited by their on-chip scratchpad memory} (SPM) \textit{arrays}. SFQ logic gates can naturally implement the gate-level pipelining, i.e., a clock pulse triggers a SFQ gate to transfer the stored SFQ to its adjacent gates. By a pulse-driven clock, SFQ circuits flow many data pulses through one wire simultaneously to achieve high operating frequency. However, SFQ-based decoders cost significant hardware overhead~\cite{Nagasawa:ASX2000,Nagasawa:TAS2001}, because the maximal fan-out of a SFQ gate is only 2~\cite{Pasandi:ISCAS2018}. Therefore, it is economical and convenient to implement shift-register-based memory (SHIFT) arrays comprising only serially-connected delay-flip-flops for a SFQ systolic CNN accelerator, since SHIFT fully utilizes the SFQ gate-level pipelining and does not require complex controls. However, SHIFT makes the SOTA SFQ systolic CNN accelerator SuperNPU~\cite{Ishida:MICRO2020} achieve only 40\% of its maximal inference throughput when processing a large batch of images, due to the lack of random access capability. Moreover, SuperNPU can only reach 16\% of its peak inference throughput when inferring a single image. Nowadays most clients are sensitive to the end-to-end latency of cloud-based services. It is more likely for data centers~\cite{Gupta:HPCA2020} to process CNN inferences with only small batch sizes, e.g., one image, simply because they are required to respond the clients rapidly and have no time to form a large batch.

It is difficult to construct a fast, dense, and power-efficient on-chip SPM architecture with random access capability for SFQ CNN accelerators by prior cryogenic memory technologies. SFQ logic works only at the 4K cryogenic temperature, so the SPM of a SFQ-based CNN accelerator has to use cryogenic memory technologies that can maintain their functionality and reliability at 4K. SOTA cryogenic memory technologies include Vortex Transition Memory (VTM)~\cite{Tahara:JAP1989,Semenov:TAS2019}, Josephson-CMOS SRAM~\cite{Masamitsu:TAS2016,Ghoshal:ISSCDTP1993,Nagasawa:TAS2001}, Magnetic Memory (MRAM)~\cite{Nguyen:SR2020}, and Superconducting Nanowire Memory (SNM)~\cite{Butters:SST2021,Zhao:SST2018}. First, prior cryogenic memory technologies use SFQ-based decoders, thereby suffering from large hardware overhead, due to the fan-out limitation of SFQ gates. Second, the scalability of VTM is poor, although accessing a VTM array costs only $\unit[0.1]{ns}$. A VTM cell~\cite{Semenov:TAS2019} is composed of four Josephson Junctions (JJs) and occupies $\unit[99]{\mu m^2}$ at the $\unit[600]{\mu A/\mu m^2}$ technology. A large capacity VTM-based SPM requires prohibitively large chip area. Third, Josephson-CMOS SRAM, MRAM, and SNM have too long access latency to match the ultra-high operating frequency of a SFQ CNN accelerator. For instance, accessing a $\unit[28]{MB}$ SRAM array at 4K requires 2$\sim$$\unit[4]{ns}$, while writing a MRAM or SNM cell costs $>$$\unit[2]{ns}$. Such long access latency seriously deteriorates the inference throughput of a SFQ CNN accelerator.

In this paper, we propose a novel heterogeneous \textbf{S}cratchpad \textbf{M}emory \textbf{AR}chi\textbf{T}ecture, \textbf{SMART}, for SFQ systolic CNN accelerators to improve their inference throughput. Our contributions are summarized as follows.

\begin{figure}[t!]
\centering
\subfigure[A SFQ ring.]{
   \includegraphics[width=1.1in]{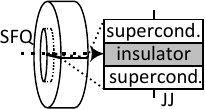}
   \label{f:sfq_jj_arch}
}
\hspace{-0.1in}
\subfigure[A SFQ DFF.]{
   \includegraphics[width=1in]{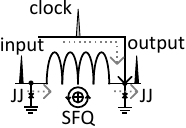}
   \label{f:sfq_jj_wave}
}
\hspace{-0.15in}
\subfigure[DFF operations.]{
   \includegraphics[width=1in]{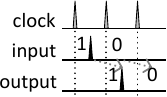}
   \label{f:sfq_jj_wave2}
}
\vspace{-0.2in}
\caption{Josephson Junction and SFQ Delay-Flip-Flop.}
\label{f:sfq_jj_all}
\vspace{-0.3in}
\end{figure}

\begin{itemize}[nosep,leftmargin=*]
\item \textbf{A comparison of cryogenic memory technologies}: We compared a variety of SFQ-compatible cryogenic memory technologies including VTM, Josephson-CMOS SRAM, MRAM, and SNM on the SOTA SFQ systolic CNN accelerator, SuperNPU. We found that no prior cryogenic memory technology can support Super-NPU to obtain high inference throughput, low power consumption, and small hardware overhead at the same time. 

\item \textbf{A heterogeneous SPM architecture}: We present a heterogeneous SPM architecture that combines SHIFT arrays and a random-access-memory (RANDOM) array to support ultra-fast sequential accesses and fast random accesses. A SFQ CNN accelerator can store its sequentially accessed data in SHIFT arrays and randomly accessed data in the RANDOM array separately.

\item \textbf{A pipelined CMOS-SFQ RANDOM array}: We propose a dense CMOS-SFQ RANDOM array for SMART to achieve fast and power-efficient random accesses. We built a pipelined SFQ-based H-Tree by SFQ passive transmission lines (PTLs) to decrease the access latency and energy consumption. Our pipelined CMOS-SFQ array uses SFQ-based H-Trees to connect CMOS sub-banks, each of which consists of SRAM cells and CMOS peripherals, e.g., row decoders, column multiplexers, and sense amplifiers.

\item \textbf{An ILP-based compiler}: We formulated the allocation and prefetching of input, weight, output, and PSum data to SMART as an integer-linear-programming (ILP) problem. Our ILP-based compiler makes near-optimal schedules for various CNN models on a SFQ systolic CNN accelerator with SMART.

\item \textbf{Inference throughput and throughput per Watt}: We evaluated and compared SMART to the SOTA SFQ systolic CNN accelerator, SuperNPU. Under the same area constraint, compared to SuperNPU, SMART improves the inference throughput by $3.9\times$ ($2.2\times$), and reduces the inference energy by $86\%$ ($71\%$) when inferring a single image (a batch of images).
\end{itemize}
The paper is organized as: SFQ logic and cryogenic memories are introduced in Section~\ref{s:back}. Section~\ref{s:moti} describes design motivation. SMRAT is proposed in Section~\ref{s:smart}. We present experiment methodology and results in Section~\ref{s:em} and Section~\ref{s:result} respectively. Related work is presented in Section~\ref{s:related}, followed by our conclusion in Section~\ref{s:con}.

\section{Background}
\label{s:back}

\subsection{SFQ Technology}

\textbf{Josephson Junction}. Superconductor SFQ logic~\cite{Tannu:CF2019,Likharev:TAS1991} is one of the most promising emerging technologies for ultra-fast and low-power computing at cryogenic temperatures. A basic element of SFQ technology, i.e., a superconductor ring~\cite{Likharev:TAS1991}, is shown in Figure~\ref{f:sfq_jj_arch}. Instead of voltage levels in CMOS logic, SFQ circuits use the existence of a single magnetic flux quantum (SFQ) in the superconductor ring to represent ``1'' or ``0''. A superconductor ring stores and transfers the SFQ by Josephson junctions (JJs)~\cite{Tolpygo:TAS2017,Tolpygo:TAS2018}, each of which consists of a thin insulator sandwiched by two superconductors. A JJ can reliably operate at $\sim\unit[70]{GHz}$. Each JJ switching costs only $\sim\unit[10^{-19}]{J}$.

\begin{figure}[t!]
\centering
\subfigure[Latency.]{
   \includegraphics[width=1.55in]{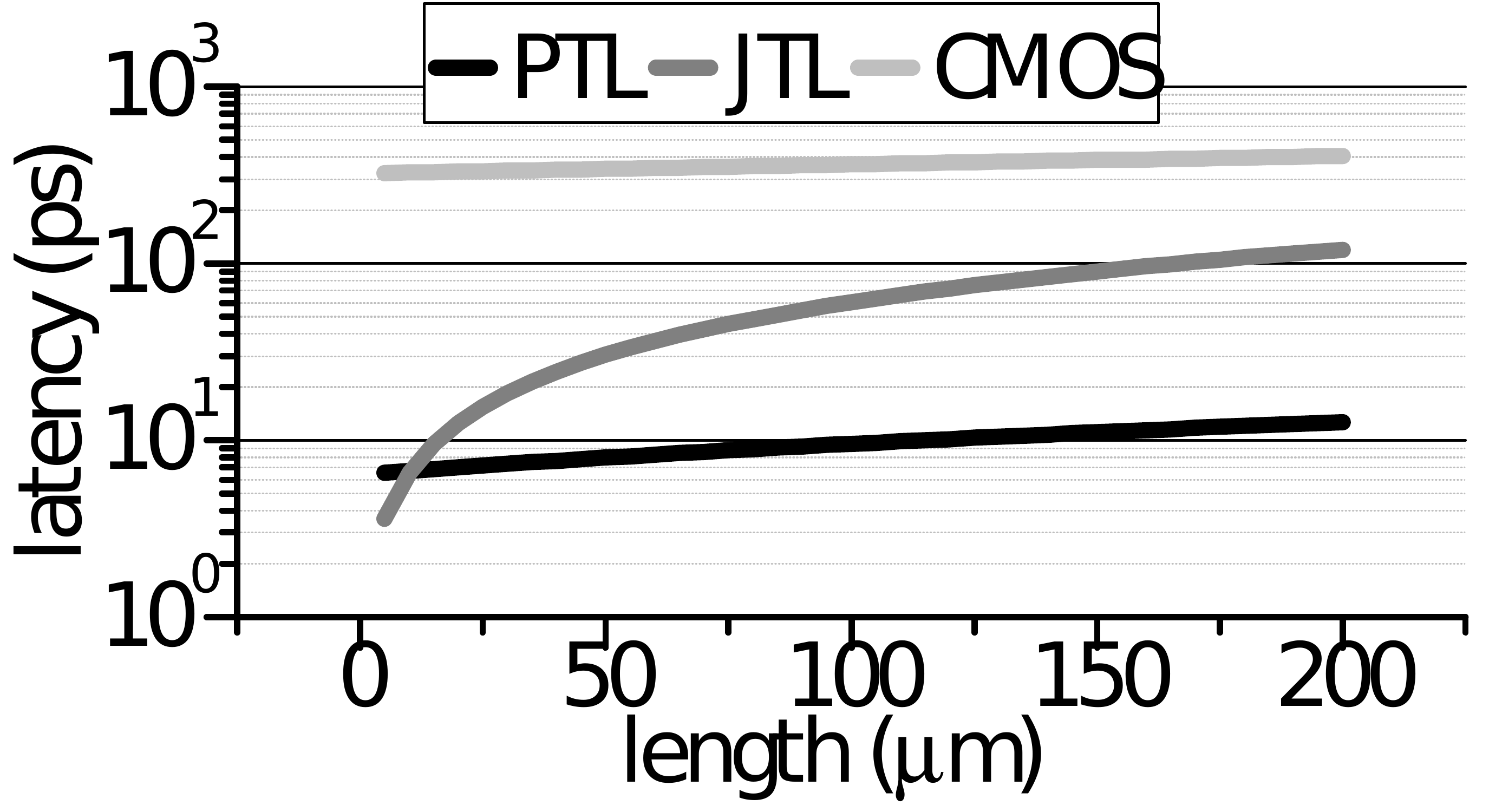}
   \label{f:sfq_line_lat}
}
\hspace{-0.1in}
\subfigure[Energy.]{
   \includegraphics[width=1.55in]{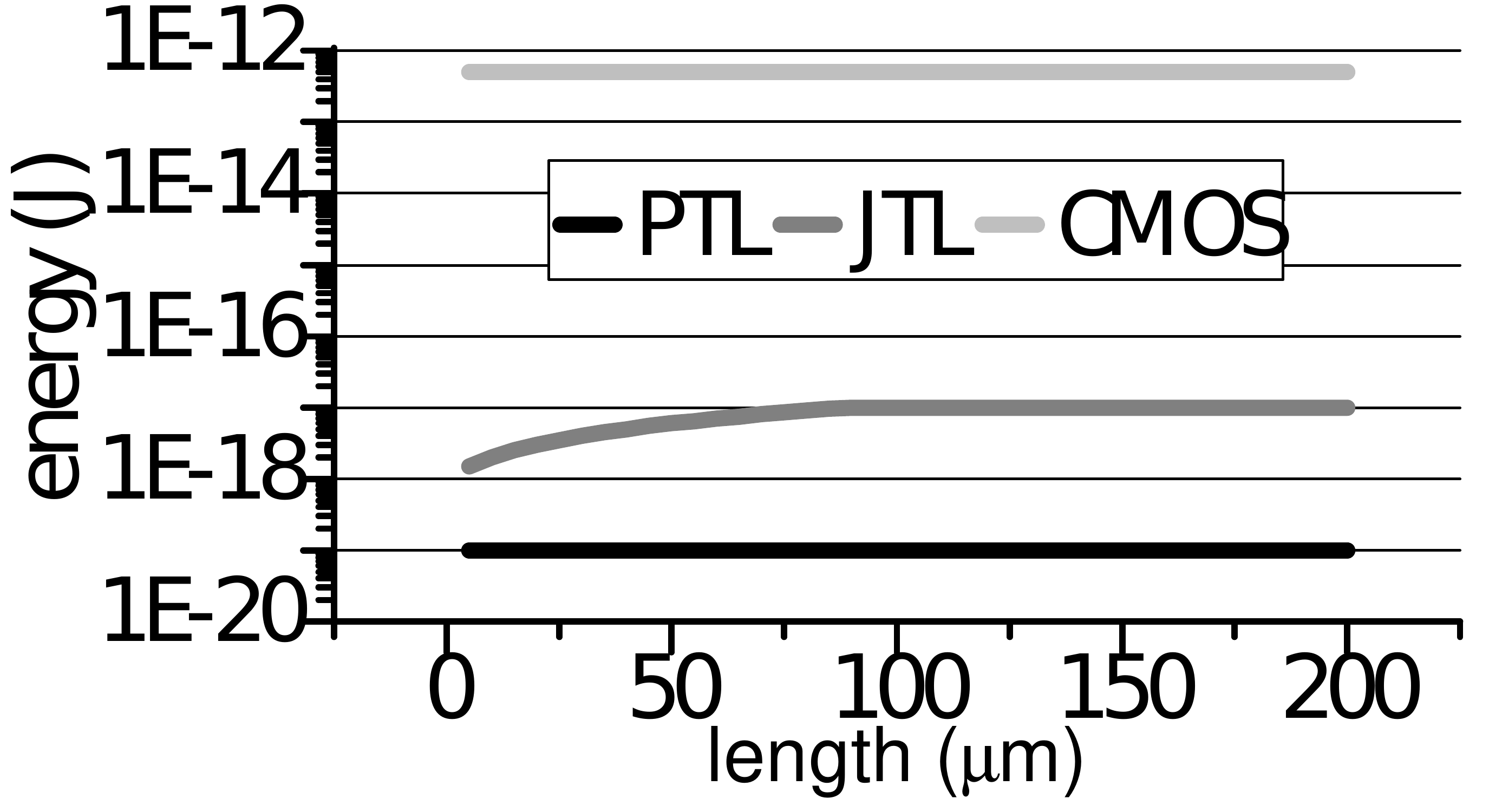}
   \label{f:sfq_line_en}
}
\vspace{-0.2in}
\caption{A comparison between SFQ and CMOS wires.} 
\label{f:sfq_line_all}
\vspace{-0.35in}
\end{figure}

\textbf{SFQ Delay-Flip-Flop}. To explain the working mechanism of SFQ logic, we use a SFQ-based delay-flip-flop (DFF) as an example because of its simple structure, i.e., it consists of only a single superconductor ring and a clock line. As Figure~\ref{f:sfq_jj_wave} shows, an input pulse makes the current flowing through the left JJ higher than its \textit{critical current} $I_c$. And then, the left JJ produces a voltage pulse, which is stored in the ring as a SFQ. When a clock pulse arrives, the right JJ is activated, and the SFQ in the ring is outputted as a voltage pulse. A SFQ DFF passes a ``1'' as the existence of the stored SFQ between two clock pulses, as shown in Figure~\ref{f:sfq_jj_wave2}. In contrast, if there is no input pulse during a clock period, no voltage pulse (``0'') is produced on the output. Several chips~\cite{Nagaoka:ISSCC2019,Nagaoka:ISEC2019} composed of SFQ logic units and memories are fabricated and demonstrated at tens of GHz.

\textbf{SFQ Interconnect}. SFQ logic components are connected by active Josephson transmission lines (JTLs) and passive transmission lines (PTLs)~\cite{Schindler:TAS2020}. As Figure~\ref{f:sfq_line_lat} shows, compared to a CMOS wire, JTL and PTL enjoy two orders of magnitude shorter latency, since they have no DC resistance \cite{JabbariTAS2020,Jabbari:WSI2020}. A PTL requires a much smaller delay than a JTL, particularly when the length is large. Furthermore, the energy comparison between CMOS and SFQ interconnects is shown in Figure~\ref{f:sfq_line_en}. The energy of a CMOS wire is roughly six orders of magnitude greater than the energy dissipated by a PTL. To implement a long line, a JTL consumes $100\times$ more energy than a PTL.

\begin{figure*}[t!]
\centering
\includegraphics[width=5.8in]{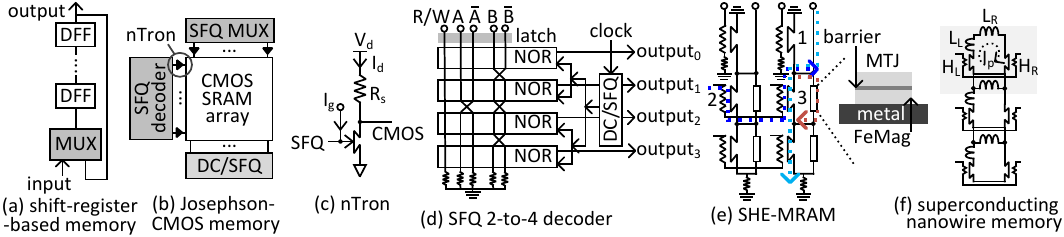}
\vspace{-0.15in}
\caption{Various cryogenic memory technologies and their components.}
\label{f:sfq_mem_all}
\vspace{-0.2in}
\end{figure*}

\textbf{SFQ Fan-out}. Unlike CMOS logic, each SFQ gate can drive only one other node~\cite{Kameda:JP2006,Pasandi:ISCAS2018}, due to the use of SFQ pulses. That is to say, the fan-out of a SFQ gate is only one. If a gate needs to have $>$$1$ fan-out, a SFQ splitter is required to be inserted at the output of the gate to enable a fan-out of two. To support additional fan-outs, a binary tree of SFQ splitters can be used. Because of the fan-out constraint, it is expensive to implement peripherals of a memory array by SFQ logic. For instance, a SFQ 4-to-16 decoder fabricated by the NEC Nb standard process occupies $\unit[885]{\mu m}$$\times$$\unit[350]{\mu m}$~\cite{Nagasawa:SST1999}, i.e., $\unit[77K]{\mathcal{F}^2}$, where we define $\mathcal{F}$ as the diameter of a JJ. However, we synthesized a $\unit[28]{nm}$ CMOS 4-to-16 decoder occupying only $\unit[18.7]{\mu m^2}$, i.e., $\unit[23K]{F^2}$, where $F$ is the technology node size, i.e., $\unit[28]{nm}$.

\textbf{CMOS Compatibility}. Superconducting SFQ technology is CMOS compatible~\cite{Potts:IEEPSMT2001}. A CMOS SRAM array and SFQ peripherals have been successfully fabricated on the same wafer~\cite{Ghoshal:ISSCDTP1993}. CMOS circuits optimized for cryogenic temperatures are first fabricated on a wafer. SFQ logic can subsequently be fabricated on the same wafer using standard SFQ process technology~\cite{Ghoshal:ISSCDTP1993}.

\begin{figure}[ht!]
\vspace{-0.1in}
\centering
\includegraphics[width=2.7in]{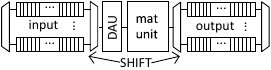}
\vspace{-0.15in}
\caption{SuperNPU: a SFQ-based systolic CNN accelerator (DAU: data alignment unit).}
\label{f:sfq_super_npu}
\vspace{-0.25in}
\end{figure}

\subsection{SuperNPU and SHIFT}

To accelerate deep learning inferences, a recent work~\cite{Ishida:MICRO2020} proposes a SFQ systolic CNN accelerator, SuperNPU, as shown in Figure~\ref{f:sfq_super_npu}. Due to the gate-level pipelining and the pulse-driven clocking, it would be easy to implement systolic and pipelined matrix multiplication units that can operate at $\unit[52.6]{GHz}$ with low power consumption by SFQ logic. Instead of power-hungry hardware-managed caches~\cite{Banakar:CODES2002}, SuperNPU uses only SHIFT~\cite{Ishida:MICRO2020} as its on-chip SPM arrays to store input, weight, output, and PSum data. As Figure~\ref{f:sfq_mem_all}(a) shows, SHIFT comprises serially connected DFFs and a feedback loop. As Table~\ref{t:sfq_mem_com} describes, due to its simple structure, SHIFT can achieve ultra-short access latency, high density, and low power consumption. An access to a SHIFT cell requires only $\unit[0.02]{ns}$ and consumes only $\unit[0.1]{fJ}$. A SHIFT cell occupies only $\unit[39]{\mathcal{F}^2}$, where $\mathcal{F}$ is the diameter of a JJ. However, SHIFT arrays seriously limit the inference throughput of SuperNPU, i.e., sequentially accessing CNN data makes SuperNPU achieve only 40\% of its peak inference throughput even when processing a batch of images.

\subsection{Cryogenic Memory}

Though SFQ-based computing logic units~\cite{Hara:TAS2009,Filippov:PP2012,Tanaka:ISSCC2004,Tomohiro:TAS2017,Kirichenko:TAS2019} achieve ultra-high operating frequency and low power consumption, it is challenging to implement low-power and dense random-access-memory (RANDOM) arrays that can match the speed of superconducting computing at 4K. There are several types of cryogenic memory technologies that can serve as on-chip SPM for a SFQ systolic CNN accelerator.

\textbf{Vortex Transition Memory (VTM)}. JJ-based Vortex Transition Memory (VTM)~\cite{Tahara:JAP1989,Semenov:TAS2019} has been demonstrated at the scale of 512-byte. However, VTM suffers from poor scalability. As Table~\ref{t:sfq_mem_com} shows, each VTM cell~\cite{Semenov:TAS2019} consists of four JJs and eight inductors, thereby occupying a cell size of $\unit[203]{\mathcal{F}^2}$. A VTM cell must use large superconductor rings. It is difficult to create a VTM cell in a smaller size even with self-shunted JJs. As a result, a recent VTM array demonstration~\cite{Semenov:TAS2019} achieves only $\unit[0.9]{Mbit/cm^2}$ functional density. Accessing a VTM array typically costs $\unit[0.1]{ns}$~\cite{Tahara:JAP1989,Semenov:TAS2019}.

\begin{table}[ht!]
\vspace{-0.1in}
\small
\setlength{\tabcolsep}{3pt}
\centering
\caption{The comparison between cryogenic memories.}
\vspace{-0.1in}
\begin{tabular}{|l||c|c|c|c|c|}
\hline
Features                   & SHIFT     & VTM           & SRAM      & MRAM     & SNM     \\\hline\hline
Read Latency (ns)          & 0.02      & 0.1           & $2\sim 4$ & 0.1      & 0.1     \\\hline
Write Latency (ns)         & 0.02      & 0.1           & $2\sim 4$ & 2        & 3       \\\hline
Cell Size                  & $39\mathcal{F}^2$ & $203\mathcal{F}^2$ & $146F^2$        & $89\mathcal{F}^2$       & $54\mathcal{F}^2$       \\\hline
Read Energy                & $0.1fJ$   & $0.1pJ$       & $0.1pJ$  & $1pJ$    & $10fJ$       \\\hline
Write Energy               & $0.1fJ$   & $0.1pJ$       & $0.1pJ$  & $8pJ$    & $10fJ$       \\\hline
Leakage Power              & no        & tiny          & medium     & tiny     & tiny       \\\hline
Random Access              & no        & yes           & yes        & yes      & yes         \\\hline
\end{tabular}
\label{t:sfq_mem_com}
\vspace{-0.1in}
\end{table}

\begin{figure*}[t!]
\centering
\begin{minipage}{.7\textwidth}
\centering
\subfigure[Latency.]{
   \includegraphics[width=1.5in]{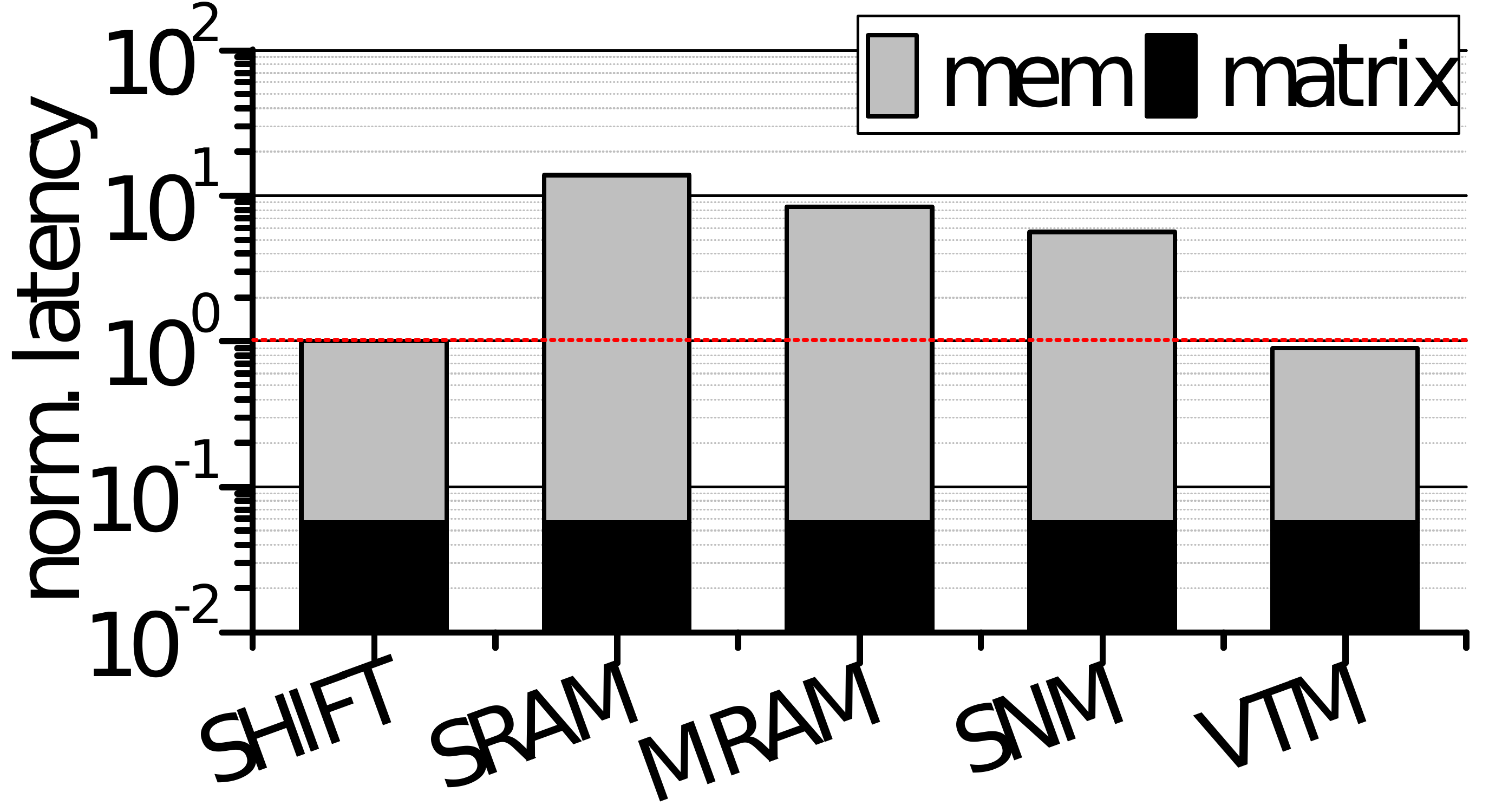}
   \label{f:sfq_mem_perf}
}
\hspace{-0.1in}
\subfigure[Energy.]{
   \includegraphics[width=1.5in]{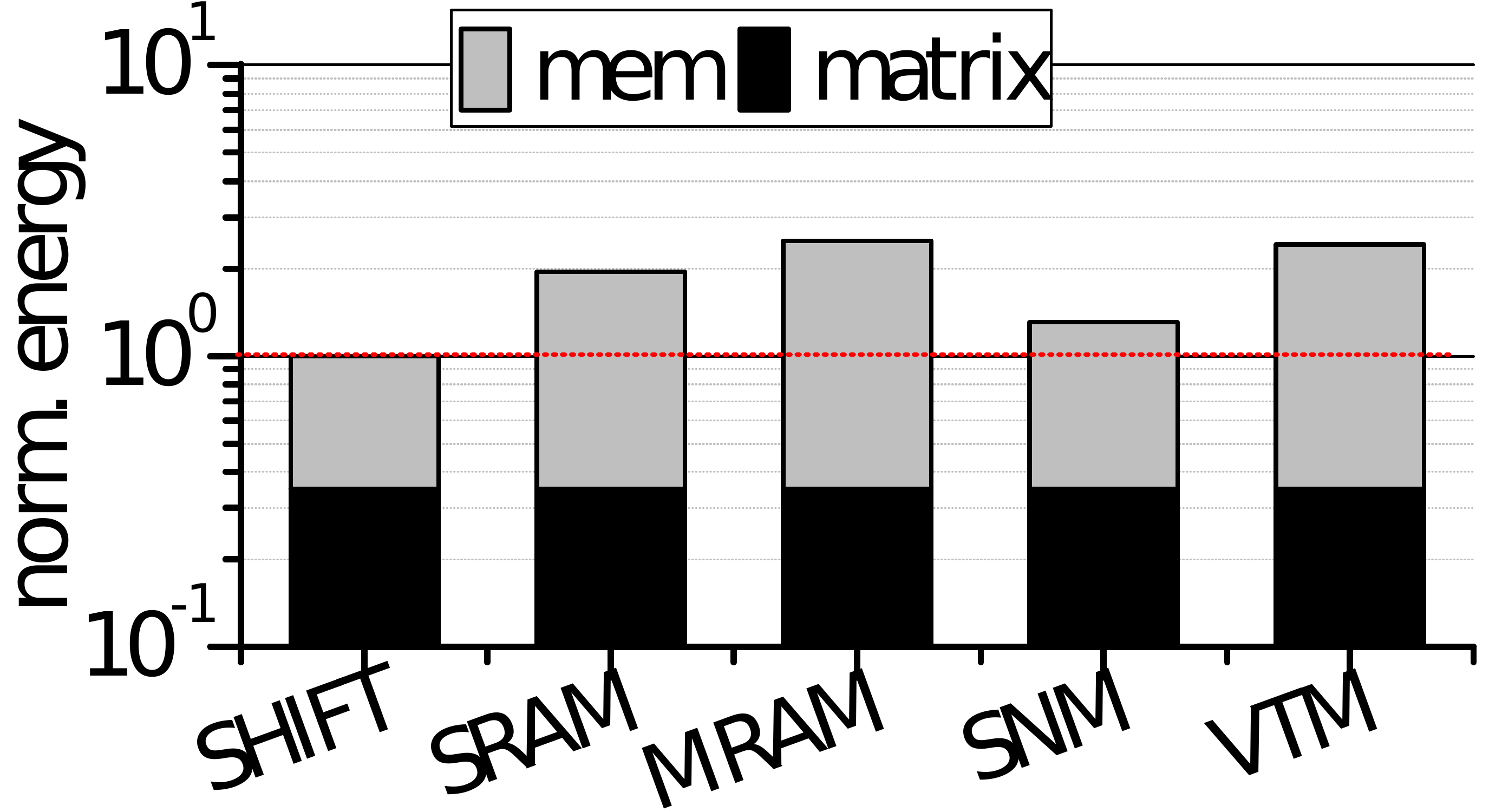}
   \label{f:sfq_mem_energy}
}
\hspace{-0.1in}
\subfigure[Area ($@28nm$).]{
   \includegraphics[width=1.5in]{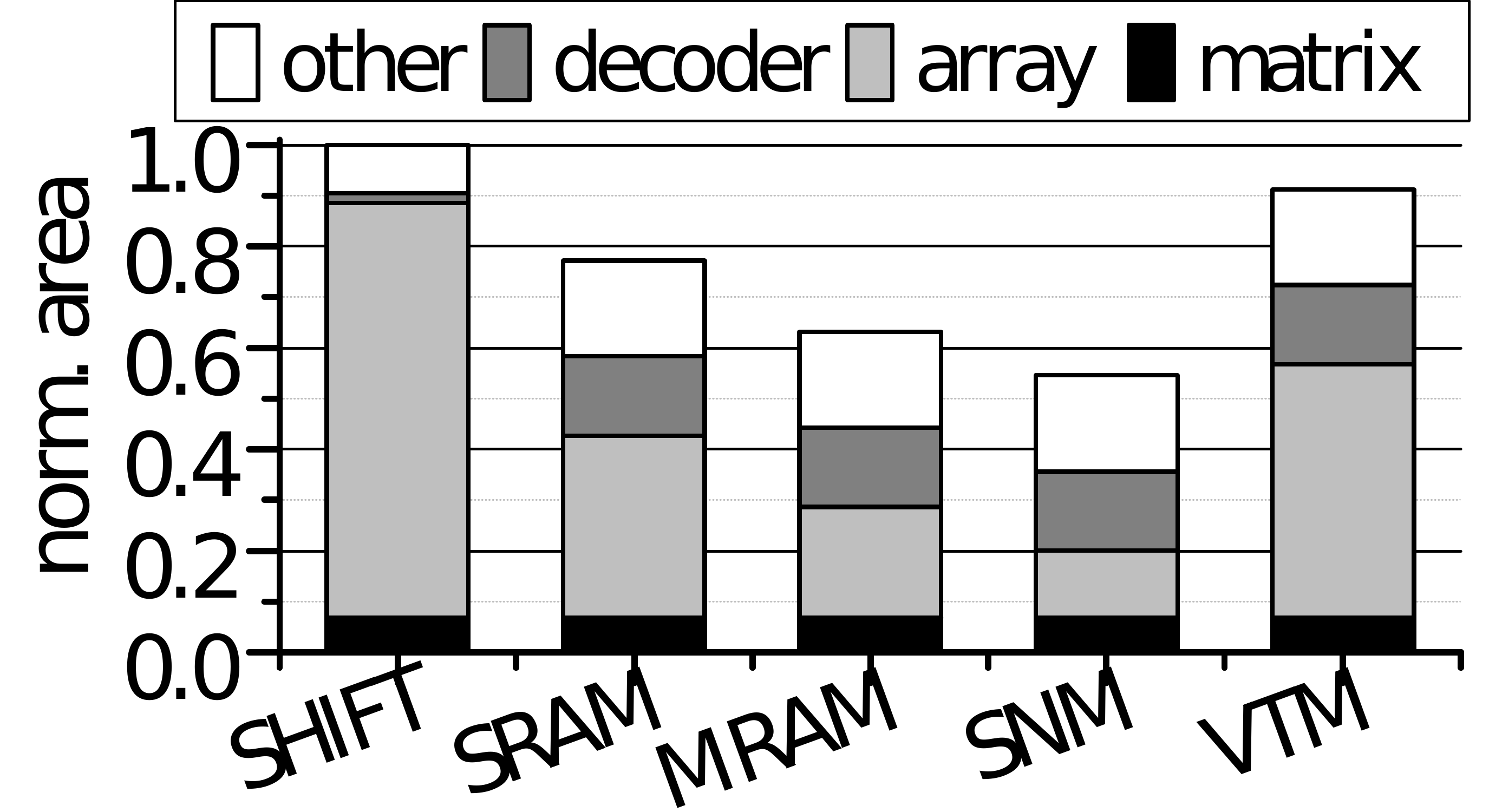}
   \label{f:sfq_mem_area}
}
\vspace{-0.2in}
\caption{The comparison of SuperNPU with various cryogenic-memory-techno-\\logy-based SPM when inferring AlexNet (mem: SPM; matrix: matrix unit).}
\label{f:sfq_mem_all0}
\end{minipage}%
\hspace{0.05in}
\begin{minipage}{0.25\textwidth}
\centering
\includegraphics[width=1.6in]{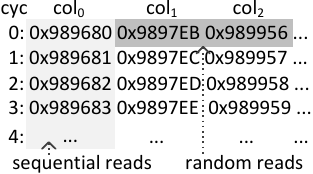}
\vspace{-0.15in}
\caption{Memory accesses of SuperNPU (cyc: cycle; col: PE array column).}
\label{f:sfq_mem_trace}
\end{minipage}
\vspace{-0.2in}
\end{figure*}

\textbf{Josephson-CMOS SRAM}. Due to the SFQ CMOS compatibility, prior work~\cite{Masamitsu:TAS2016,Ghoshal:ISSCDTP1993,Nagasawa:TAS2001,VanDuzer:TAS2013} builds a Josephson-CMOS memory array that connects a SFQ decoder and a SFQ multiplexer to a SRAM array via nTrons~\cite{Zhao:SST2017}, as shown in Figure~\ref{f:sfq_mem_all}(b). These works~\cite{Masamitsu:TAS2016,Ghoshal:ISSCDTP1993,Nagasawa:TAS2001,VanDuzer:TAS2013} have demonstrated that SRAM can reliably operate at 4K but with faster speed and lower power consumption compared to the room temperature. As Figure~\ref{f:sfq_mem_all}(c) highlights, nTron is a superconducting device whose superconductivity can be switched by the injection of hot quasiparticles generated at the gate. SFQ circuits can use nTrons to access CMOS components at $\unit[10]{GHz}$~\cite{Zhao:SST2017}. Therefore, it is more practical to implement large and reliable cryogenic memory arrays by Josephson-CMOS SRAM, due to the maturity of CMOS SRAM technology. However, it is important to note that SRAM is slow, e.g., accessing a $\unit[28]{MB}$ SRAM array typically costs $2$$\sim$$\unit[4]{ns}$, as shown in Table~\ref{t:sfq_mem_com}. Moreover, a SFQ-based decoder~\cite{Nagasawa:TAS2001} costs significant hardware overhead. Due to the fan-out limitation, as shown in Figure~\ref{f:sfq_mem_all}(d), a SFQ-based $N$-to-$2^N$ decoder requires at least $\mathcal{O}(2^N)$ SFQ splitters to distribute its clock pulses. A SFQ decoder~\cite{Nagasawa:SST1999} is larger than its CMOS counterpart by multiple times, even if JJ can be scaled to the same size of a transistor.

\textbf{Magnetic Memory (MRAM)}. To build a fast, dense, and power-efficient cryogenic memory array, recent work~\cite{Nguyen:SR2020} suggests a spin hall effect (SHE) magnetic RAM (MRAM) array, as shown in Figure~\ref{f:sfq_mem_all}(e). A SHE-MRAM cell consists of a SHE magnetic tunnel junction (MTJ) and a superconducting heater-cryotron (hTron) bit-select element. A SHE-MTJ consists of a MTJ sitting on a metallic spin hall channel, while a hTron, which is a variant of the nTron, can be driven by SFQ logic and thus supports sufficient current to switch the SHE-MTJ. A SHE-MRAM cell is $\unit[89]{\mathcal{F}^2}$, as shown in Table~\ref{t:sfq_mem_com}. Besides SFQ decoders and multiplexers, a SHE-MTJ array is connected to row and column driving hTrons. To write a cell, the SFQ multiplexer sends a triggering pulse to each corresponding column hTron. The bias current (1) flows through all hTrons in the column, which are superconducting. A row hTron is triggered by the SFQ decoder and sends its bias current to all bit-select hTrons in that row (2). For a hTron which receives both the current from the column driver and the current from the row driver, a writing pulse is generated to the SHE-MTJ channel to change the state of the MTJ (3). The switching of a SHE-MRAM typically costs $\unit[2]{ns}$~\cite{Nguyen:SR2020}. The reading process is similar to that of writing, except that the reading current is much smaller.

\textbf{Superconducting Nanowire Memory (SNM)}. A Superconducting Nanowire Memory (SNM)~\cite{Butters:SST2021,Zhao:SST2018} can be also used to build a cryogenic memory array. As Figure~\ref{f:sfq_mem_all}(f) shows, each SNM cell has two hTrons, such that the right hTron has a larger switching current and larger inductance than the left hTron. The two hTrons are connected serially so that both hTrons are modulated by the same current. The cell has four connections arranged in two electrically isolated pairs, wherein one is the access port, while the other is the select port. As Table~\ref{t:sfq_mem_com} shows, a SNM cell is only $\unit[54]{\mathcal{F}^2}$. To write a cell, a bias current is applied to the column, and flows through all the cells within the column, but its amplitude is too small to alter the state of any cells. A row enabling current is applied to the row. This weakens the channels of the hTrons within the row, thereby allowing the write bias to cause the selected cells to switch. A write operation spends $\unit[3]{ns}$~\cite{Butters:SST2021,Zhao:SST2018}. Each read is destructive. After each read, a write operation is required to restore the data.

\vspace{-0.1in}
\section{Motivation}
\label{s:moti}

In this section, we present the design motivation by comparing the inference latency, energy consumption, and area overhead of SuperNPU~\cite{Ishida:MICRO2020} with SPMs made by various cryogenic memory technologies. SuperNPU has two $\unit[24]{MB}$ SHIFT-based SPMs for inputs and outputs/PSums, respectively. We used other cryogenic memory technologies that support random accesses to build a 64-bank $\unit[12]{MB}$ input SPM, a 256-bank $\unit[16]{MB}$ output/PSum SPM, and a $\unit[64]{KB}$ weight SPM for SuperNPU. We evaluated SuperNPU for one-image inferences, thus SPMs with such capacities are large enough for each layer of AlexNet without generating thrashing traffic to DRAM. The configuration of SuperNPU is shown in Section~\ref{s:em}.

\textbf{Inference Latency}. As Figure~\ref{f:sfq_mem_perf} shows, SuperNPU using SH-IFT spends a huge portion of inference latency in sequentially searching the input and PSum data. If SuperNPU SPMs support random accesses, the inference latency can be reduced. However, since Jose-phson-CMOS SRAM, VTM, MRAM, and SNM have much longer read and write latencies, no prior cryogenic memory technology can significantly reduce the inference latency. The write latencies of SRAM, MRAM, and SNM are $>$$\unit[2]{ns}$, they prolong the inference latency of SuperNPU by at least $5\times$. Only VTM decreases the inference latency of SuperNPU by 11\% over SHIFT, since the latency saving introduced by its random access capability is larger than the slowdown caused by its prolonged access latency. If there were a random access array with $\unit[0.02]{ns}$ latency, SuperNPU would have eliminated memory access stalls. Such fast random access arrays can reduce the inference latency of SuperNPU by 94\%.

\textbf{Inference Energy}. The energy comparison of various types of on-chip SPM arrays is shown in Figure~\ref{f:sfq_mem_energy}. Since all the other cryogenic memory technologies have larger read and write energy than SHIFT, they enlarge the energy of an AlexNet inference by $30\%$$\sim$$2.5\times$ over SHIFT. Although CMOS SRAM dissipates large leakage power at room temperatures, the cryogenic temperatures substantially reduce leakage by $>$$90\%$~\cite{Min:ASPLOS2020}. As a result, the large write energy makes cryogenic SHE-MRAM consume even more energy than Josephson-CMOS SRAM.

\begin{figure*}[t!]
\centering
\begin{minipage}{.23\textwidth}
\centering
\includegraphics[width=1.6in]{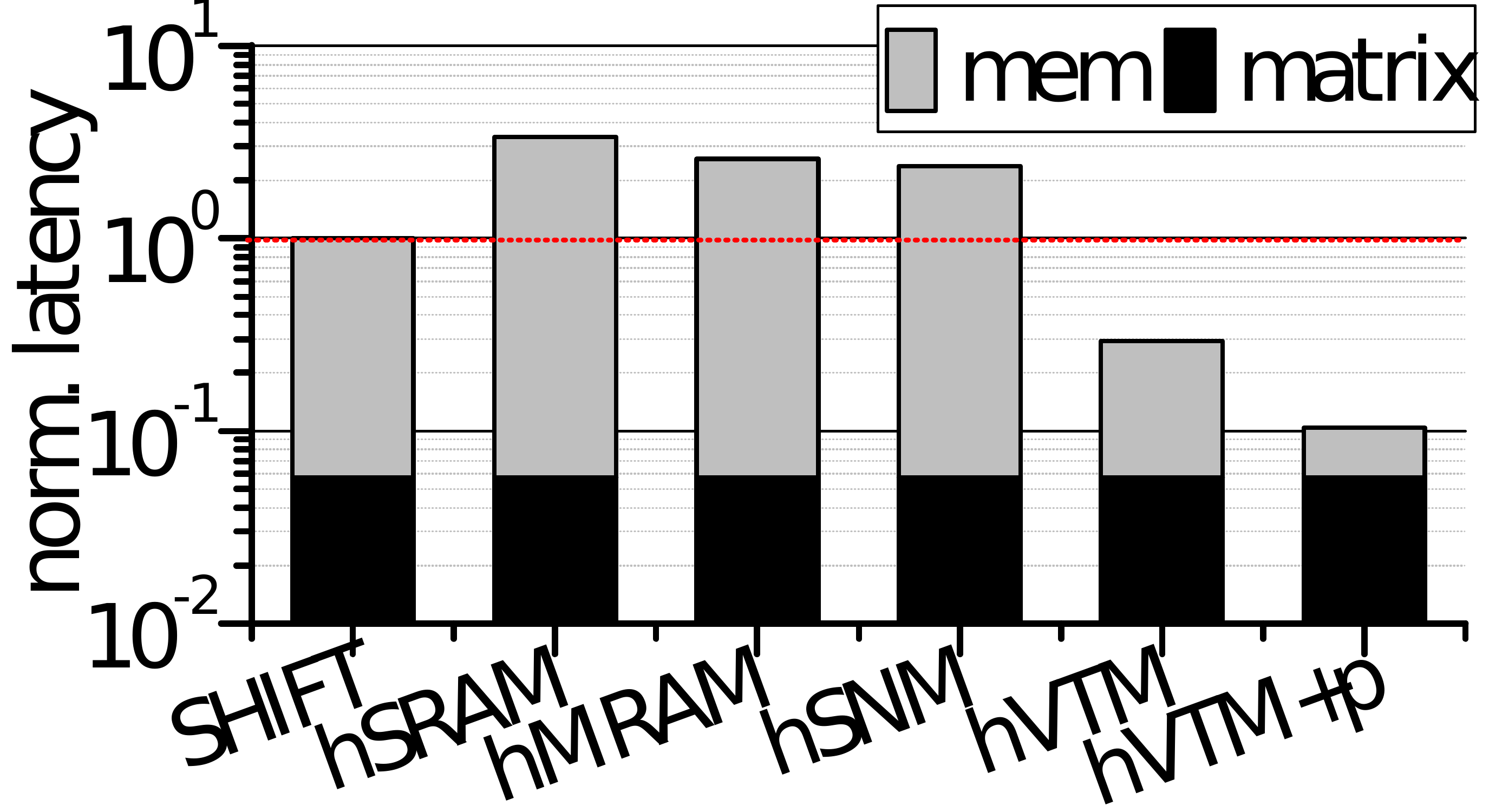}
\vspace{-0.3in}
\caption{The inference laten-cy comparison of a heterogeneous SPM.}
\label{f:sfq_hybrid_perf0}
\end{minipage}
\hspace{0.1in}
\begin{minipage}{.23\textwidth}
\centering
\includegraphics[width=1.5in]{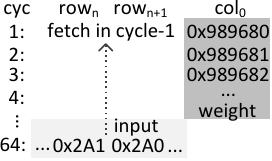}
\vspace{-0.15in}
\caption{SMART prefetching (cyc: cycle; row: PE array row; col: PE array column).}
\label{f:sfq_mem_prefetch}
\end{minipage}%
\hspace{0.1in}
\begin{minipage}{.23\textwidth}
\centering
\includegraphics[width=1.5in]{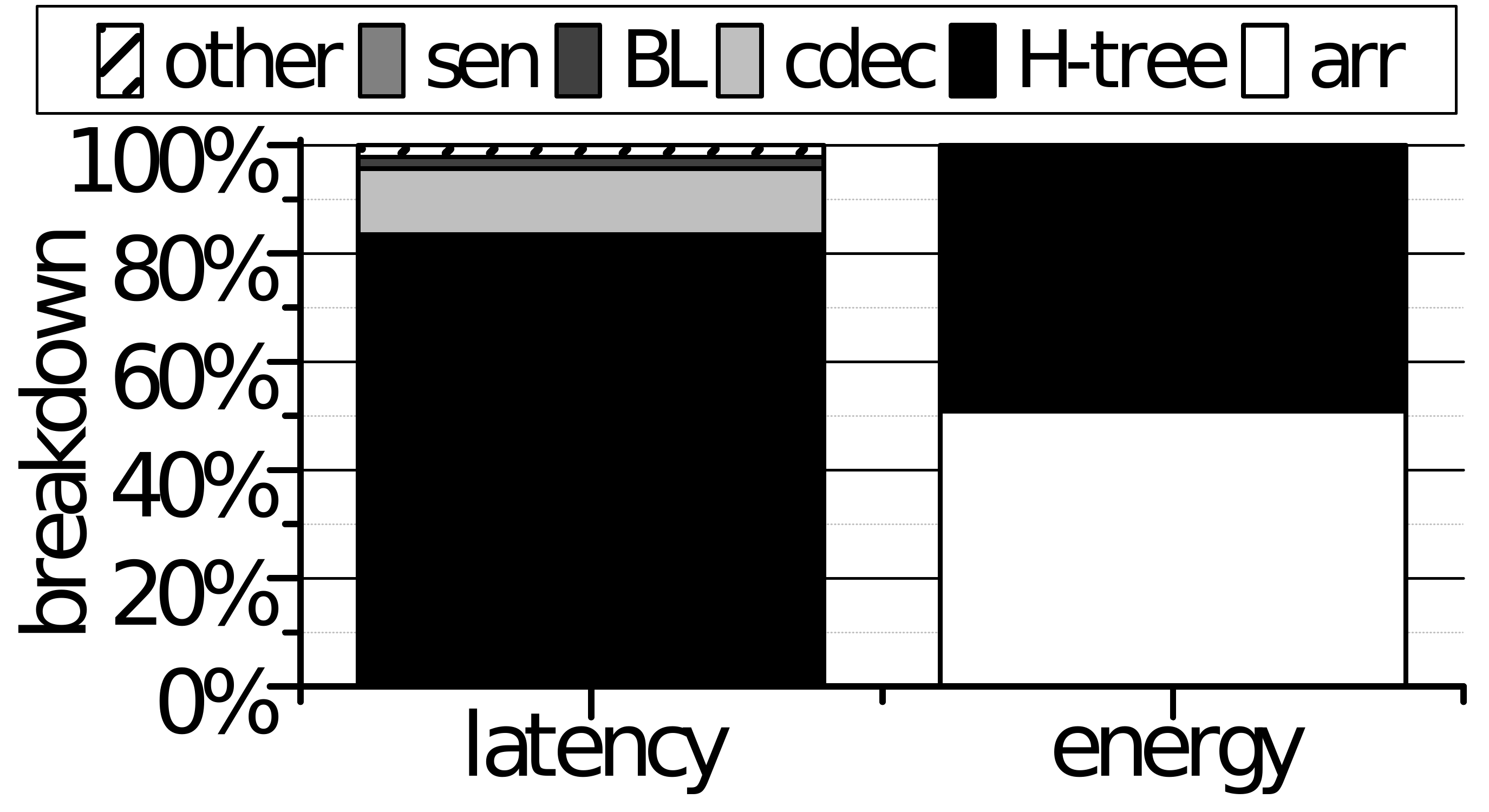}
\vspace{-0.15in}
\caption{The latency \& ener-gy of CMOS H-Trees in $\unit[28]{MB}$ Josephson-CMOS array with 256 banks.}
\label{f:sfq_sram_all9}
\end{minipage}%
\hspace{0.1in}
\begin{minipage}{.24\textwidth}
\centering
\includegraphics[width=1.3in]{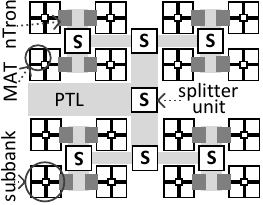}
\vspace{-0.1in}
\caption{A CMOS-SFQ array.}
\label{f:sfq_htree_arch}
\end{minipage}
\vspace{-0.2in}
\end{figure*}

\textbf{Area Overhead}. The area comparison between various types of on-chip SPM arrays is highlighted in Figure~\ref{f:sfq_mem_area}. SuperNPU~\cite{Ishida:MICRO2020} assumes JJs can be scaled to $\unit[28]{nm}$. We adopted the same assumption for SHIFT-, MRAM-, SNM-, and VTM-based SPM arrays. We also assumed SRAM arrays are fabricated at $\unit[28]{nm}$. The SHIFT SPMs of SuperNPU have few SFQ decoders and multiplexers to select banks, each of which is a long lane of SHIFT memory cells. Although the capacity of MRAM-, SNM-, and VTM-based SPM arrays is $58\%$ of that of SHIFT, they can reduce from $8\%$ to $45\%$ of the area. This is because they use more SFQ peripherals and have larger cells, which are demonstrated in Table~\ref{t:sfq_mem_com}. Particularly, SFQ-based decoders cost $16\%$$\sim28\%$ of the area in non-SHIFT arrays. Due to the fact that Josephson-CMOS SRAM has the second largest cell size, compared to SHIFT, the Josephson-CMOS SRAM array with a $58\%$ capacity reduces the area by only $22\%$.

\textbf{Drawbacks of Prior Cryogenic Memories}. Compared to the perfect pipeline without memory stall, the SHIFT-based SPMs prolong the inference latency of SuperNPU by $17\times$, due to the fact that it only supports sequential reads. As the memory traces in Figure~\ref{f:sfq_mem_trace} show, when SuperNPU reads weights, it has both sequential and random reads. Although SHIFT-based SPM can efficiently process sequential reads, it also has to move many unnecessary bits to support random accesses. Josephson-CMOS-SRAM-, MRAM-, SNM-, and VTM-based SPM arrays can perform random accesses, but they cannot achieve reasonable latency reduction, since their read or/and write latency are too long. MRAM and SNM are bottlenecked by their write latency and energy. Despite that VTM has the shortest access latency among prior cryogenic memory technologies, it is still not fast enough to make an observable latency reduction. Furthermore, the large VTM cell size significantly enlarges the array area. Thus, although the SFQ peripherals of Josephson-CMOS SRAM are very fast, CMOS H-Trees~\cite{Min:ASPLOS2020} inside SRAM arrays greatly degrade the access latency and energy. The area efficiency of Josephson-CMOS-SRAM-, MRAM-, SNM-, and VTM-based SPM arrays are limited by SFQ peripherals. In summary, no prior cryogenic memory technology is a good candidate to implement on-chip SPMs for SuperNPU.

\section{SMART}
\label{s:smart}

In this section, we propose a heterogeneous SPM architecture, SM-ART, in order to reduce the inference latency of a SFQ systolic CNN accelerator. SMART is composed of SHIFT arrays performing sequential accesses and a random-access-memory (RANDOM) array supporting random accesses. We further present a fast RANDOM array, i.e., a pipelined SFQ-CMOS array, for SMART to minimize the inference latency, energy and hardware area. A pipelined SFQ-CMOS array uses SFQ PTLs and splitter units to implement H-trees connecting CMOS sub-banks to achieve small access latency and energy. At last, we propose an ILP-based compiler to deploy various CNN models on SMART.

\subsection{A Heterogeneous SPM Architecture}

We present a heterogeneous SPM architecture consisting of SHIFT arrays and a RANDOM array for a SFQ systolic CNN accelerator. For each convolutional layer, SHIFT arrays store all data receiving sequential accesses, while the RANDOM array is used to support random accesses during an inference. There are two challenges we face when trying to use this heterogeneous SPM architecture to effectively reduce the inference latency of the SFQ systolic accelerator. First, though SHIFT arrays process sequential accesses well, the inference latency of the accelerator is still heavily influenced by the access latency of the RANDOM array. However, it is difficult to build a fast, dense, and power-efficient RANDOM array by prior cryogenic memory technologies. Second, there is no compilation technique that can deploy a CNN and enable prefetching on the heterogeneous SPM architecture. Although data allocation to SPMs has been heavily studied before, prior work~\cite{Suhendra:RTSS2005,Udayakumaran:TECS2006,Deverge:ECRTS2007,Verma:TVLSI2006,Liu:ICCD2012} focuses only on general-purpose applications running on CPUs.

\begin{figure*}[t!]
\centering
\includegraphics[width=7in]{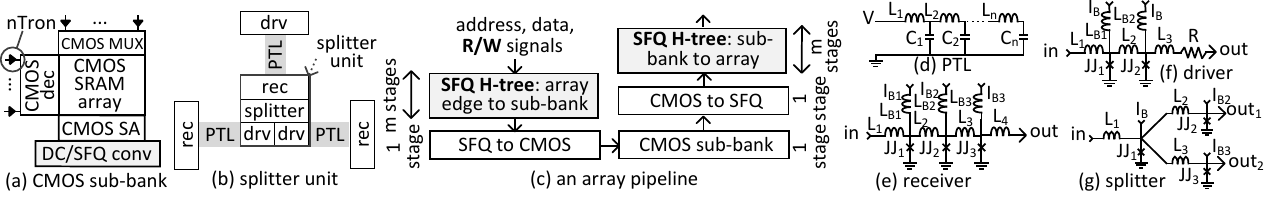}
\vspace{-0.3in}
\caption{The components and pipeline of a pipelined CMOS-SFQ array.}
\label{f:sfq_hybrid_bank}
\vspace{-0.2in}
\end{figure*}

We elaborate the two challenges in applying heterogeneous SPMs on SuperNPU in Figure~\ref{f:sfq_hybrid_perf0}, where we assume a perfect data allocation for both sequentially accessed data and randomly accessed data. We consider three $\unit[32]{KB}$ SHIFT arrays for inputs, outputs \& PSums, and weights as their SPMs, respectively. All CNN data share a $\unit[28]{MB}$ 256-bank RANDOM array in the heterogeneous SPM architecture. The RANDOM array can be built by Josephson-CMOS-SRAM, MRAM, SNM, or VTM. We call these heterogeneous SPM schemes hSRAM, hMRAM, hSNM, and hVTM in Figure~\ref{f:sfq_hybrid_perf0}. Compared to SHIFT, hSRAM, hMRAM, and hSNM prolong the inference latency by $3.36\times$, $2.59\times$, and $2.38\times$, respectively. hVTM reduces the inference latency by 70\% over SHIFT, due to its short access latency. We find that the RANDOM array access latency in SMART heavily influences the inference latency of the accelerator. This is because for a weight-stationary systolic CNN accelerator, most accesses to input, and output \& PSum data are random. The systolic accelerator maintains an iterative computing flow, where weights are first deployed on the matrix unit, inputs are fetched to start a systolic computation, and then the next iteration continues, as shown in Figure~\ref{f:sfq_mem_prefetch}. Considering the fact that there is no dependency between inputs and weights, if the prefetching of inputs to its SPM is enabled, we can start the systolic computation earlier. As Figure~\ref{f:sfq_hybrid_perf0} shows, the prefetching (hVTM+p) further reduces the inference latency by 64.4\% over hVTM. However, no prior SPM management technique supports prefetching for an accelerator.

\subsection{A Pipelined CMOS-SFQ Array}

\subsubsection{The limitations imposed by CMOS H-trees}

In an array, both the address and data of a memory request are routed by H-Trees~\cite{Muralimanohar:MICRO2007}, which make the memory request consistent in its access to all MATs. A memory array has two separate H-Trees including a request network and a reply network. Data and addresses are transferred from the edge of the array to MATs by the request network, while data are sent out from MATs by the reply network. Both the request and reply H-Trees are composed of two parts including a network connecting the array edge to the bank edge, and a network connecting the bank edge to MATs.

The Josephson-CMOS array access latency can be divided into SFQ decoder delay, CMOS H-Tree delay, CMOS decoder delay, CMOS wordline delay, CMOS bitline delay, CMOS sense amplifier delay, and SFQ DC/SFQ delay. Throughout the components, the CMOS H-tree dominates the latency and energy consumption of a large Josephson-CMOS SRAM array at 4K. As Figure~\ref{f:sfq_sram_all9} shows, the H-tree costs $84\%$ of the access latency, and $49\%$ of the access energy in a 256-bank $\unit[28]{MB}$ Josephson-CMOS SRAM array. Particularly, in the sub-$10nm$ regime, the resistance of copper wires~\cite{Chen:SLIP2012} exponentially increases as the process technology scales. Therefore, the latency and energy consumption of H-trees will become more significant in Josephson-CMOS arrays at future process nodes.

\subsubsection{A Pipelined CMOS-SFQ Array}
\label{s:pipeline}

\textbf{Overall Architecture}. We propose a pipelined CMOS-SFQ array as shown in Figure~\ref{f:sfq_htree_arch} to reduce the access latency and energy at 4K. Our pipelined CMOS-SFQ array consists of only CMOS sub-banks connected by SFQ H-Trees. The design philosophy of our CMOS-SFQ array is different from Josephson-CMOS SRAM~\cite{Masamitsu:TAS2016,Ghoshal:ISSCDTP1993,Nagasawa:TAS2001}. To avoid the large hardware overhead of SFQ decoders, we use SRAM cells and CMOS peripherals including row decoders, column multiplexers, and sense amplifiers. We use PTL lines and SFQ-based peripherals including splitters, drivers, receivers, and nTrons to build SFQ H-Trees. The major components of our pipelined CMOS-SFQ array can be summarized as follows.
\begin{itemize}[nosep,leftmargin=*]
\item \textbf{CMOS Sub-bank}: As Figure~\ref{f:sfq_hybrid_bank}(a) shows, CMOS sub-banks of a pipelined CMOS-SFQ array are constructed by SRAM cells and CMOS peripherals including CMOS row decoders, column multiplexers, and sense amplifiers. To drive the row decoders and column multiplexers, we use nTron devices to convert the SFQ memory requests to electrical signals for a CMOS sub-bank. After a CMOS sub-bank makes the data ready, we also use level-driven DC/SFQ converters~\cite{Masamitsu:TAS2016} to transform the data in sense amplifiers into SFQ pulses.

\item \textbf{SFQ H-Tree}: We use PTL lines to replace all CMOS (e.g., copper) lines in a pipelined CMOS-SFQ array. Due to the fan-out limitation of SFQ logic, we add a splitter unit to each position where the fan-out needs to be increased. The details of a splitter unit can be viewed in Figure~\ref{f:sfq_hybrid_bank}(b). In order to pass a SFQ pulse via a PTL line, we need a driver at the source end and a receiver at the destination end of the PTL line. A splitter unit consists of a receiver at the input end, two drivers at the two output ends, and a splitter connecting them together.
\end{itemize}

\begin{figure}[t!]
\centering
\subfigure[Latency.]{
   \includegraphics[width=1.51in]{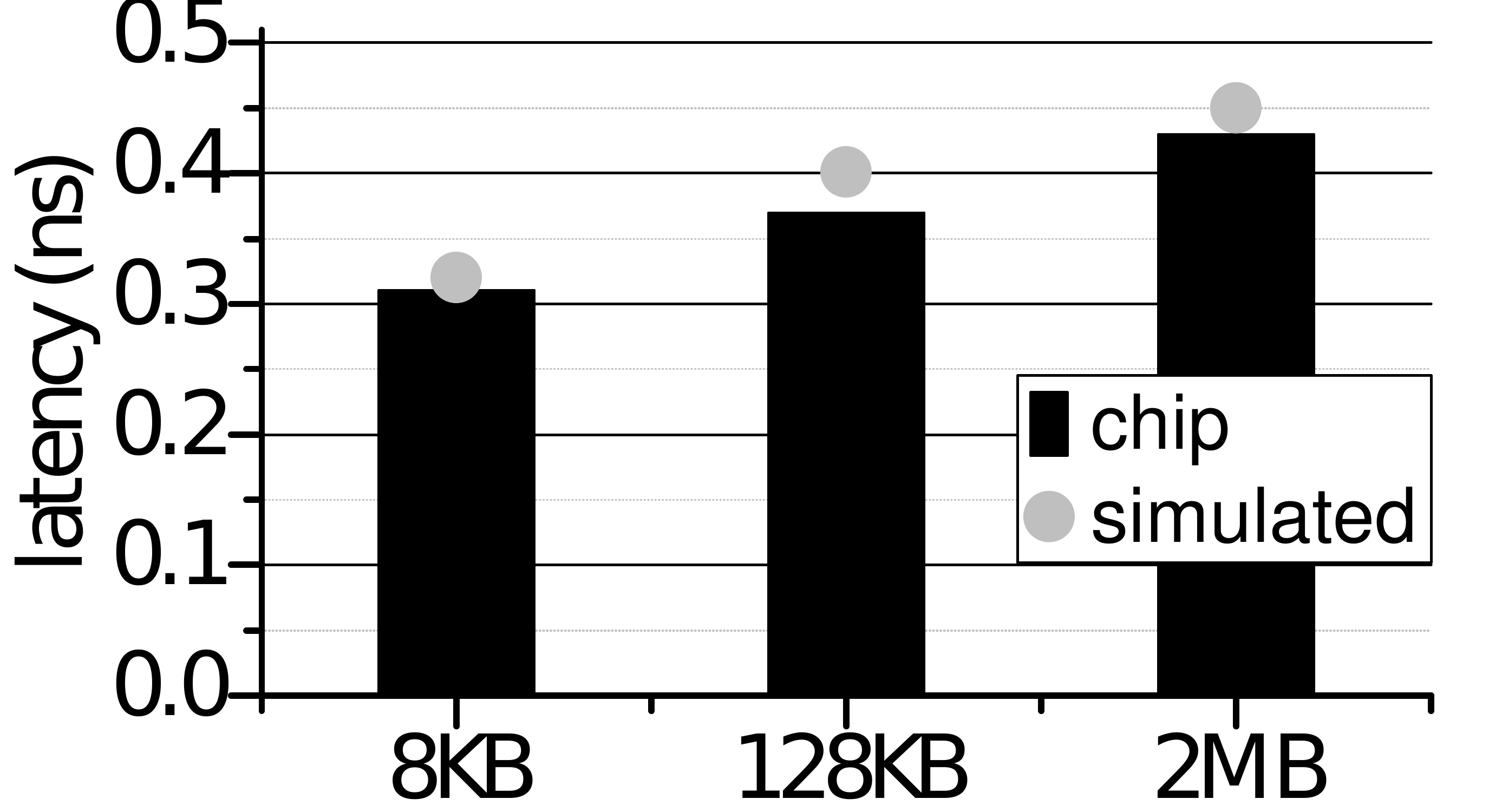}
   \label{f:sfq_bank_latency}
}
\subfigure[Energy.]{
   \includegraphics[width=1.51in]{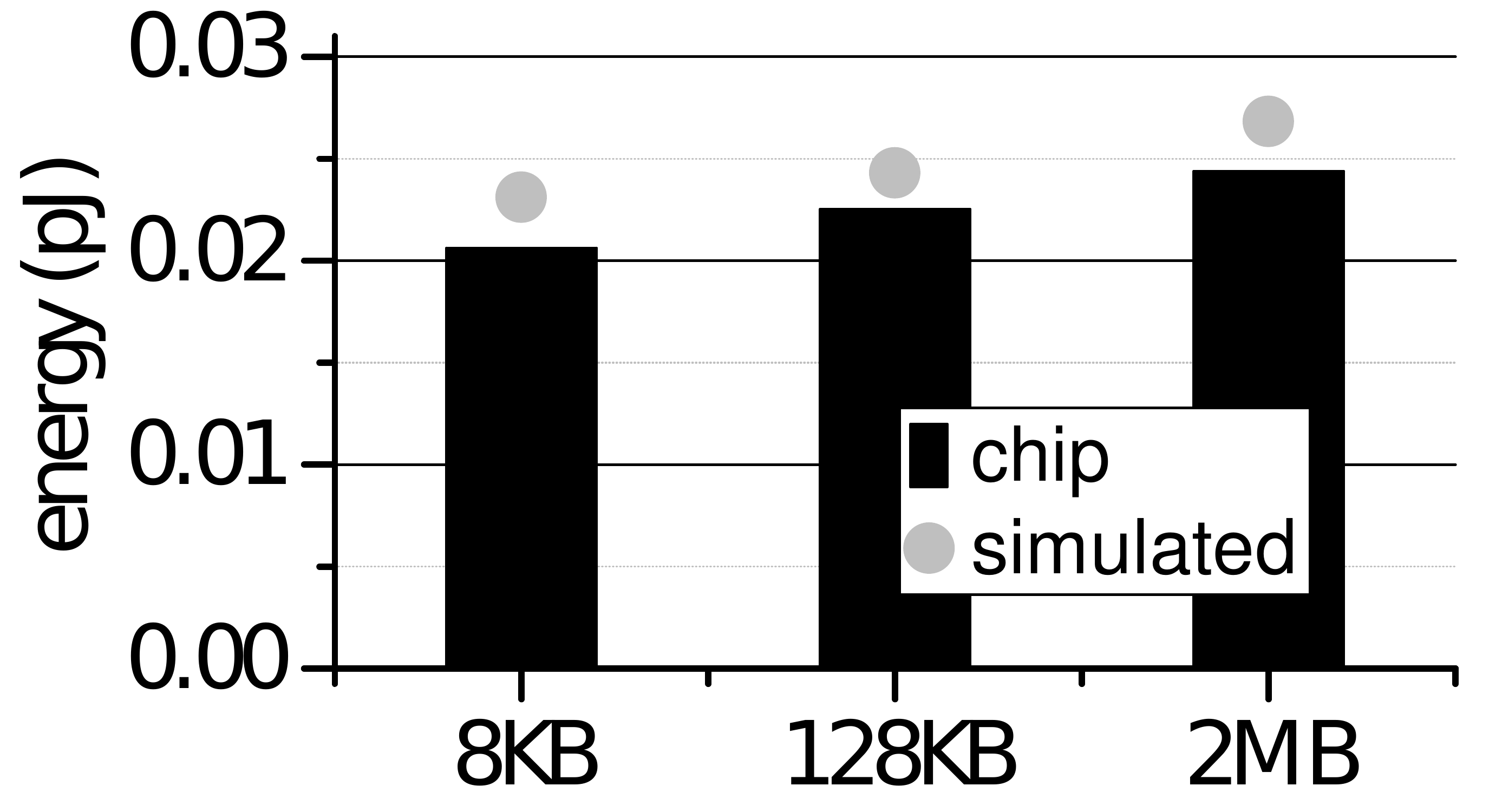}
   \label{f:sfq_bank_energy}
}
\vspace{-0.2in}
\caption{The validation of a CMOS sub-bank.}
\label{f:sfq_bank_all}
\vspace{-0.3in}
\end{figure}

\textbf{Pipeline}. We propose a multi-stage pipeline architecture for our CMOS-SFQ array in Figure~\ref{f:sfq_hybrid_bank}(c). To communicate with the SFQ systolic matrix unit, request SFQ H-trees transfer each memory request to a sub-bank from the array edge. nTrons are used to convert the SFQ request to electrical signals that can drive CMOS arrays to fetch (write) the data from (to) the CMOS sub-bank. If the request is a read, level-driven DC/SFQ converters are adopted to convert the electrical signals of the reading data back to SFQ pulses. Finally, the SFQ data pulses are returned to the systolic matrix unit via reply SFQ H-trees. Since splitter units in SFQ H-Trees naturally have gate-level pipelining, multiple memory requests can be transferred simultaneously in the same H-Tree. If we can guarantee all requests go to different sub-banks, a CMOS-SFQ array can process these requests in a pipelined way. To decide the frequency of the pipeline, we identified the operations of nTrons (SFQ to CMOS), CMOS sub-banks, and level-driven DC/SFQ converters as the bottlenecks. Both a nTron and a level-driven DC/SFQ converter~\cite{Masamitsu:TAS2016} can complete a conversion around $\unit[0.1]{ns}$. We can limit the latency of each sub-bank within $\sim$$\unit[0.1]{ns}$ by adjusting the number of MATs inside a sub-bank. Then, a H-Tree operation can be broken into multiple pipeline stages by inserting SFQ repeaters, each of which is composed of a driver and a receiver, so that each pipeline stage of H-tree can also fit into $\sim$$\unit[0.1]{ns}$. The detailed pipeline design space exploration is shown in Section~\ref{s:explore}. Since all memory accesses of a systolic CNN accelerator can be known before executions, it is possible to read (write) a line from (to) a pipelined SFQ-CMOS array every $\sim$$\unit[0.1]{ns}$ via data allocation and prefetching.

\subsubsection{Modeling and Validation}

\textbf{Modeling a CMOS Sub-bank at 4K}. We adopted the cryogenic memory model, CryoRAM~\cite{Lee:ISCA2019} to model a CMOS SRAM sub-bank. CryoRAM includes a  validated cryogenic MOSFET model cryo-pgen, and a CACTI-based cryogenic memory model cryo-mem. Cryo-pgen can derive a variety of MOSFET characteristics at only 77K. We modified cryo-pgen to model MOSFET at 4K by adjusting three fabrication-related and temperature-dependent MOSFET variables including carrier mobility, carrier's saturation velocity, and threshold voltage based on recent cryogenic MOSFET data~\cite{Beckers:IJEDS2020,Grill:IRPS2020}. Then, we plugged the 4K MOSFET parameters generated by cryo-pgen into cryo-mem to study the access latency and energy of a CMOS array at 4K.

\textbf{Validating the 4K CMOS Sub-bank Model}. We validated the access latency and energy of a CMOS array at 4K generated by cryo-mem against a published 4K SRAM array demonstration~\cite{Masamitsu:TAS2016} fabricated at $\unit[0.18]{\mu m}$. As Figure~\ref{f:sfq_bank_all} shows, the 4K SRAM demonstration has three configurations: an $\unit[8]{KB}$ sub-bank consisting of eight MATs, a $\unit[128]{KB}$ sub-bank containing 32 MATs, and a $\unit[2]{MB}$ sub-bank comprised of 128 MATs. The latency values simulated by our modified cryo-mem are larger than those of the 4K SRAM chip by $3\%$$\sim$$8\%$ as shown in Figure~\ref{f:sfq_bank_latency}, since we applied conservative cryogenic MOSFET parameters to cryo-mem. Our conservative cryogenic MOSFET parameters also make the energy values of our modified cryo-mem larger than those of the 4K SRAM chip by $8\%$$\sim$$12\%$.

\begin{table}[t!]
\small
\setlength{\tabcolsep}{3pt}
\centering
\caption{The latency and power of SFQ H-Trees.}
\vspace{-0.1in}
\begin{tabular}{|l||c|c|c|}
\hline
\multirow{2}{*}{Component}       & Latency         & Leakage            & Dynamic   \\
                                 & (ps)            & Power ($\mu W$)    & Power ($nW$) \\\hline\hline
Splitter                         & 7               & 0                  & 0.15  \\\hline
Driver                           & 3.5             & 0.874              & 0.181      \\\hline
Receiver                         & 5.25            & 0                  & 0.275      \\\hline
nTron                            & 103.02          & 8.8                & 13    \\\hline
\end{tabular}
\label{t:sfq_htree_com}
\vspace{-0.15in}
\end{table}

\begin{figure}[ht!]
\centering
\subfigure[Latency.]{
   \includegraphics[width=1.51in]{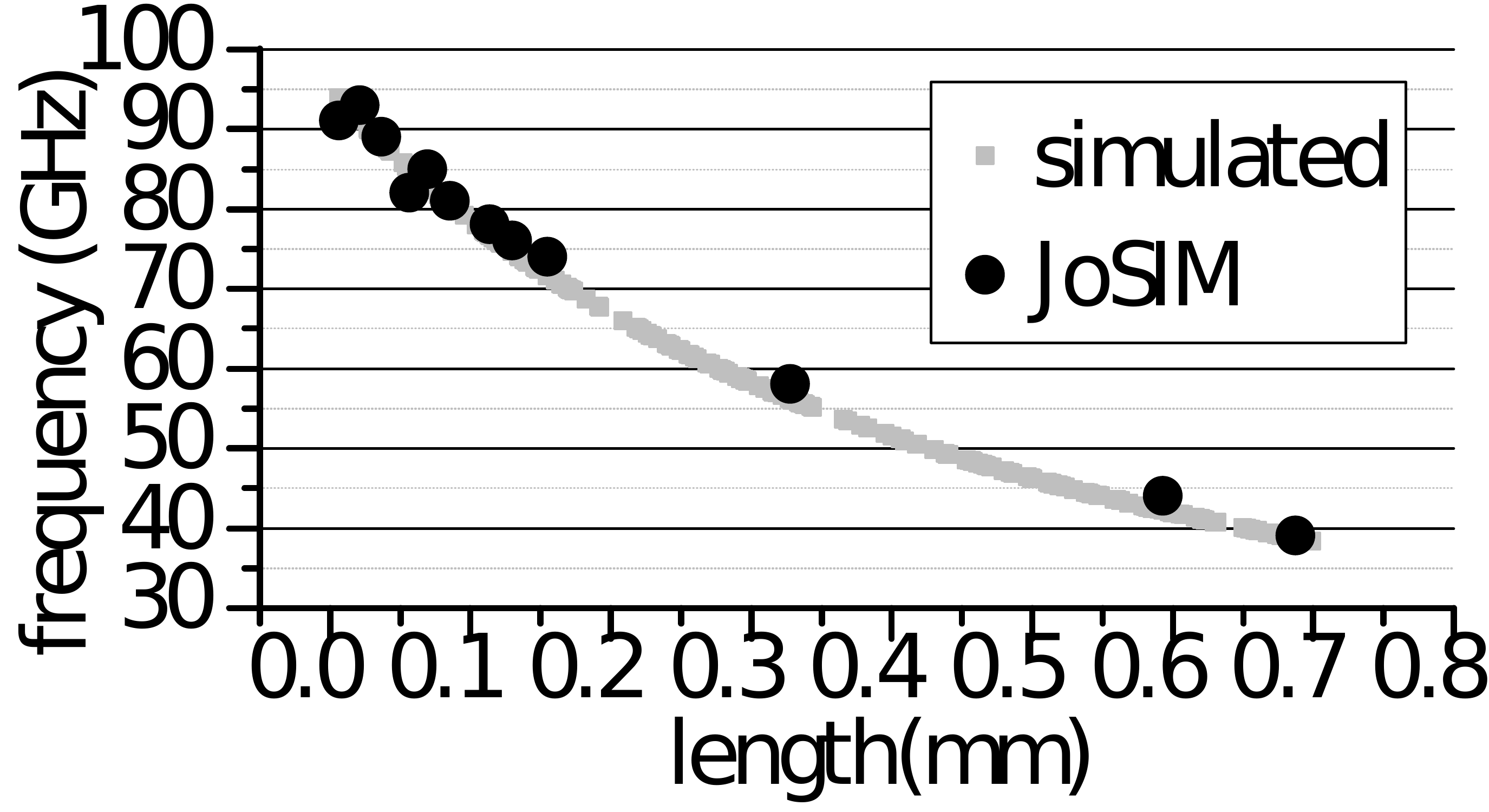}
   \label{f:sfq_model_latency}
}
\hspace{-0.1in}
\subfigure[Energy.]{
   \includegraphics[width=1.51in]{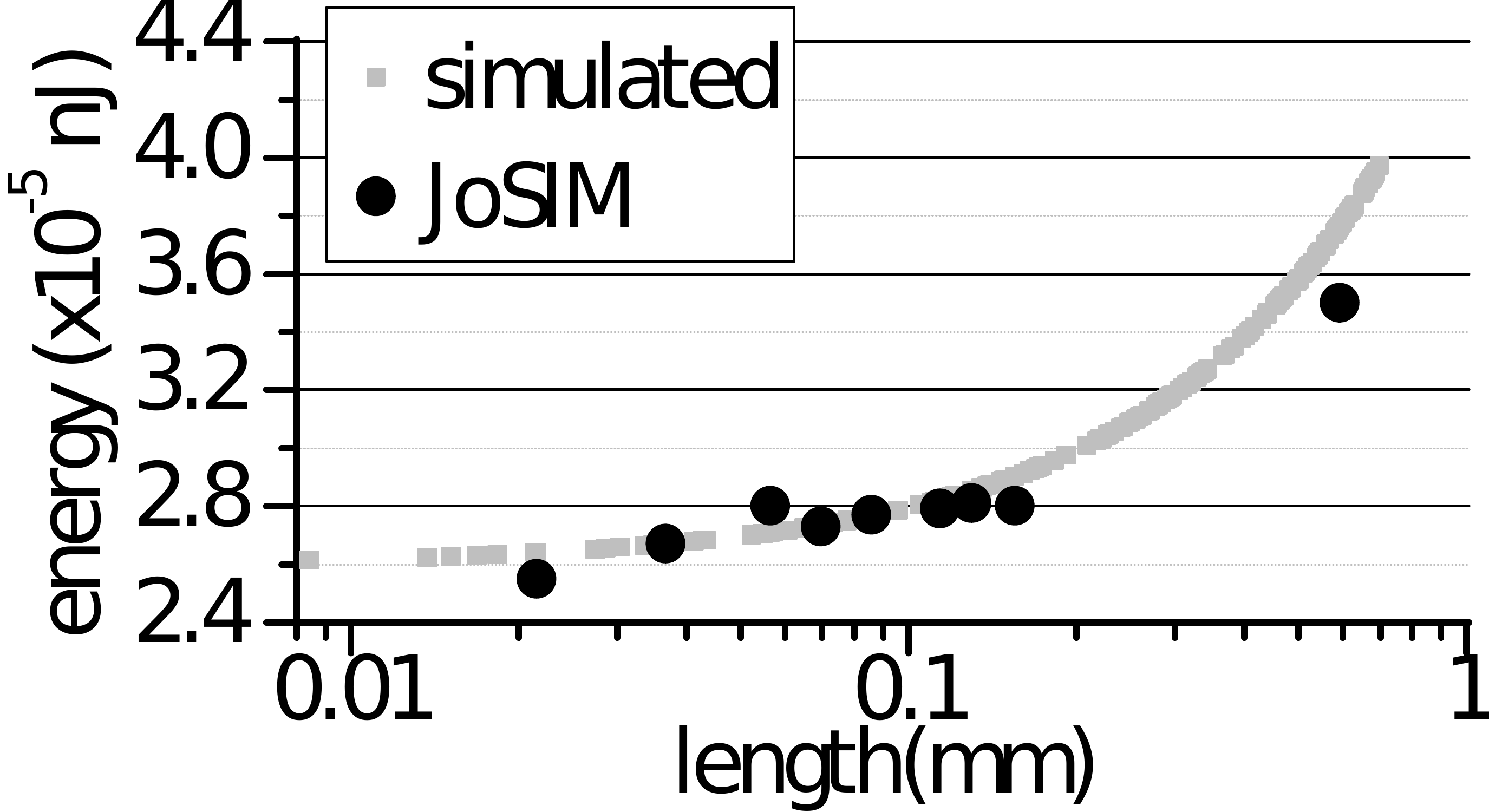}
   \label{f:sfq_model_energy}
}
\vspace{-0.2in}
\caption{The validation of our SFQ H-Tree model.}
\label{f:sfq_design_space}
\vspace{-0.2in}
\end{figure}

\textbf{Modeling a SFQ H-Tree at 4K}. The components of a SFQ H-Tree include the follows.
\begin{itemize}[nosep,leftmargin=*]
\item \textbf{PTL}: We used micro-strip PTLs~\cite{Jabbari:TAS2020}, due to its small size, better scalability and simplicity of geometry. A micro-strip PTL can be represented as a lossless distributed LC network shown in Figure~\ref{f:sfq_hybrid_bank}(d). The inductance per unit length of a micro-strip PTL ($L$)~\cite{Mohebbi:SST2009} is composed of the magnetic inductance introduced by magnetic fluxes within a superconductive line, and the kinetic inductance caused by the motion of paired electrons. $L$ can be calculated as:
\begin{equation}
	L = \frac{\mu _0 h}{K w} \left[1+\frac{\lambda _1}{h} \textrm{coth} \left(\frac{t_1}{\lambda_1}\right)+ \frac{\lambda_2}{h}  \textrm{coth}\left(\frac{t_2}{\lambda_2}\right) \right] 
\label{e:sfq_inductance}
\end{equation} 
where $w$ is the line width; $t_1$ means the thickness of the PTL; $t_2$ is the thickness of the ground plane of the PTL; $K$ indicates the fringing field factor; $h$ is the thickness of dielectric; $\lambda_1$ and $\lambda_2$ denote penetration depths of the micro-strip and the ground plane, respectively.

\begin{minipage}{.15\textwidth}
\centering
\begin{equation}
 C = \frac{\epsilon _r \epsilon_0 w}{h}
\label{e:sfq_capacitance}
\end{equation}
\vspace{0.0001in}
\end{minipage}
\begin{minipage}{.15\textwidth}
\centering
\begin{equation}
Z = \sqrt{\frac{L}{C}}
\label{e:sfq_impedance}
\end{equation}
\vspace{0.0002in}
\end{minipage}
\begin{minipage}{.15\textwidth}
\centering
\begin{equation}
T = N\sqrt{L \times C }
\label{e:sfq_latency}
\end{equation}
\vspace{0.0001in}
\end{minipage}
The capacitance per unit length of a micro-strip PTL ($C$) can be calculated by Equation~\ref{e:sfq_capacitance}, where $w$ and $h$ are defined in Equation~\ref{e:sfq_inductance}; $\epsilon_r$ is the dielectric constant of the insulation between the line and ground plane layer; and $\epsilon_0$ is the permittivity of free space. As Equation~\ref{e:sfq_impedance} shows, the impedance of a micro-strip PTL can be derived from the inductance and capacitance per unit. The delay of a micro-strip PTL is a function of total LC, and increases linearly with the line length as shown in Equation~\ref{e:sfq_latency}, where $N$ is the number of LC sections in the micro-strip PTL.

\item \textbf{Splitter}: Due to the fan-out limitation, a splitter~\cite{Pasandi:ISCAS2018} is the core of a splitter unit used to transform a pulse to two pulses, each of which can be sent in one direction of a cross-point in the H-Tree. The structure of a splitter is shown in Figure~\ref{f:sfq_hybrid_bank}(g), where a SFQ pulse is converted into two flux quanta. A splitter consists of three inductors and three JJs. The latency, and dynamic power of a splitter are shown in Table~\ref{t:sfq_htree_com}.

\item \textbf{Driver \& Receiver}: As Figure~\ref{f:sfq_hybrid_bank}(b) shows, a SFQ pulse is sent to a PTL by a driver~\cite{Schindler:TAS2020} and received by a receiver~\cite{Schindler:TAS2020}. A PTL driver in Figure~\ref{f:sfq_hybrid_bank}(f) consists of a 2-stage JTL cascaded with a resistance. The JTL acts as both a buffer and a SFQ pulse reconstruction device. A receiver composed of a 3-stage JTL is exhibited in Figure~\ref{f:sfq_hybrid_bank}(e). The resonance frequency $f$ of a PTL with a driver and a receiver is defined as $f=\frac{1}{2T+t_{0}}$, where $T$ is the PTL delay, another $T$ avoids the resonance, and $t_{0}$ is the delay of a driver and a receiver~\cite{Chonigman:TAS2021}. The operating frequency of a PTL can be set to at most 90\% of $f$~\cite{Joukov:TAS2000}. Otherwise, the resonance effect on the PTL may cause timing jitters and errors. In order to increase the frequency of a PTL, we need to insert more repeaters, each of which consists of a driver and a receiver. Therefore, a long PTL can be partitioned into shorter segments. Inserting repeaters into a PTL increases not only the resonance frequency, but also the power consumption of the PTL. The bias currents and resistors in the bias network of a driver increase the static power, while more JJs introduced by repeater insertion also increase the dynamic power. The area overhead of repeater insertion is proportional to the number of JJs.
\end{itemize}

\textbf{Validating the 4K SFQ H-Tree Model}. We implemented our pipelined SFQ H-Trees (Equation~\ref{e:sfq_inductance}$\sim$Equation~\ref{e:sfq_latency}) in the CACTI-based cryogenic memory model cryo-mem~\cite{Lee:ISCA2019}. We mainly focus on validating the new modules added to cryo-mem including PTL lines and splitter units, each of which consists of a driver, a receiver, and a splitter. Thus, we used a splitter unit shown in Figure~\ref{f:sfq_hybrid_bank}(b) with various PTL lengths to perform the validation. We measured the latency and energy of passing a SFQ pulse from the top driver to the bottom right receiver, since the two bottom receivers are the same. We ran the superconductor SPICE simulator, JoSIM~\cite{Delport:JOSIM2018}, to validate the results of pipelined CMOS-SFQ arrays generated by our modified cryo-mem. We assumed Hypres ERSFQ $1.0 \mu m$ technology~\cite{Yohannes:TAS2015} to validate the splitter unit. Figure~\ref{f:sfq_model_latency} exhibits the latency comparison of a splitter unit with various PTL lengths between our model and JoSIM, while their energy correlation is described in Figure~\ref{f:sfq_model_energy}. Compared to the JoSIM HSPICE results, the latency values of a driver and a receiver estimated by our SFQ H-Tree model have $\pm6\%$ deviations, particularly when the PTL length is $<$$0.2mm$ . The energy values of a SFQ H-Tree predicted by our model are also close to the JoSIM results with $\pm11\%$ errors.

\begin{figure}[t!]
\centering
\subfigure[Latency.]{
   \includegraphics[width=1.05in]{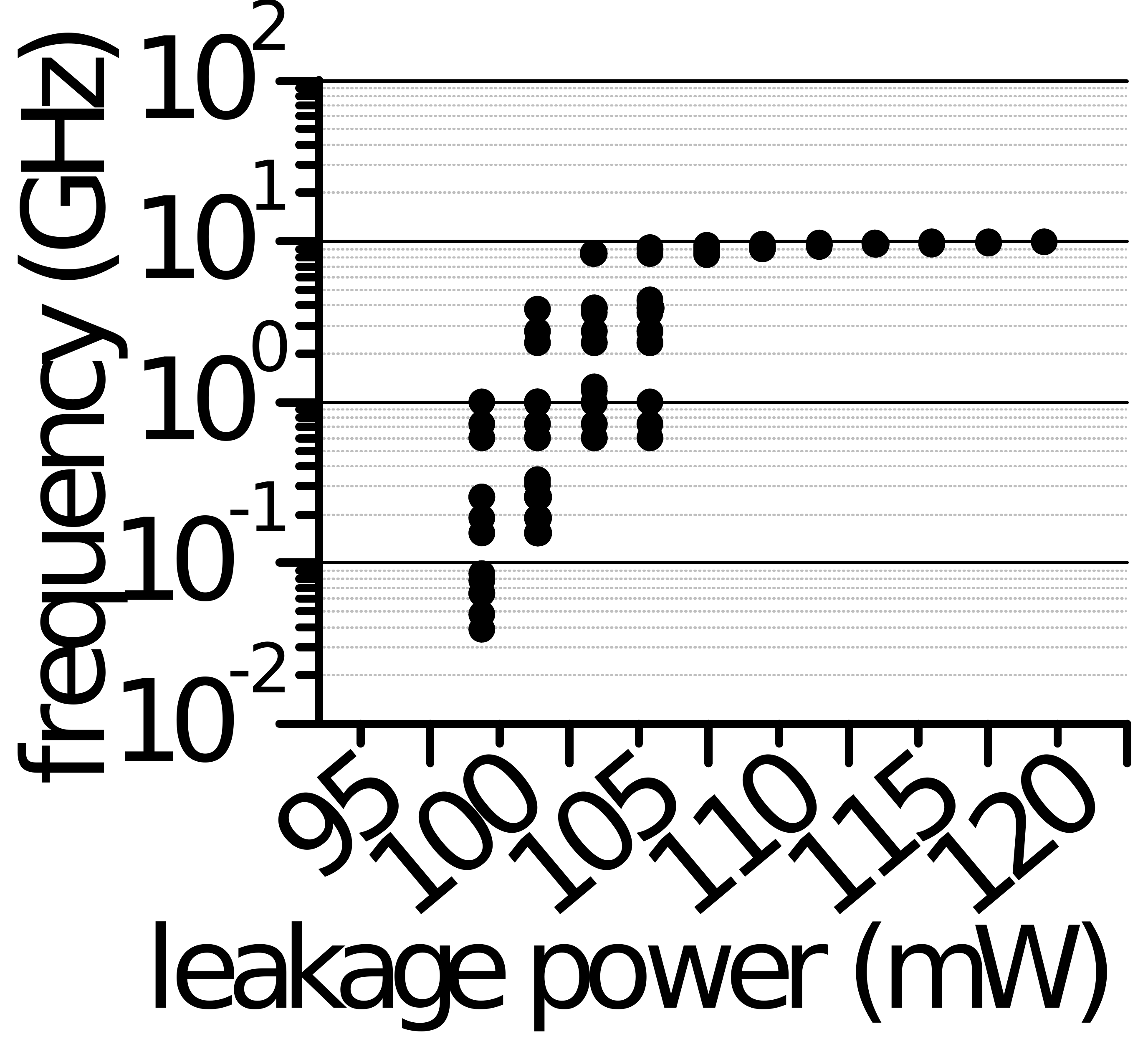}
   \label{f:sfq_latency_leakage}
}
\hspace{-0.15in}
\subfigure[Energy.]{
   \includegraphics[width=1.05in]{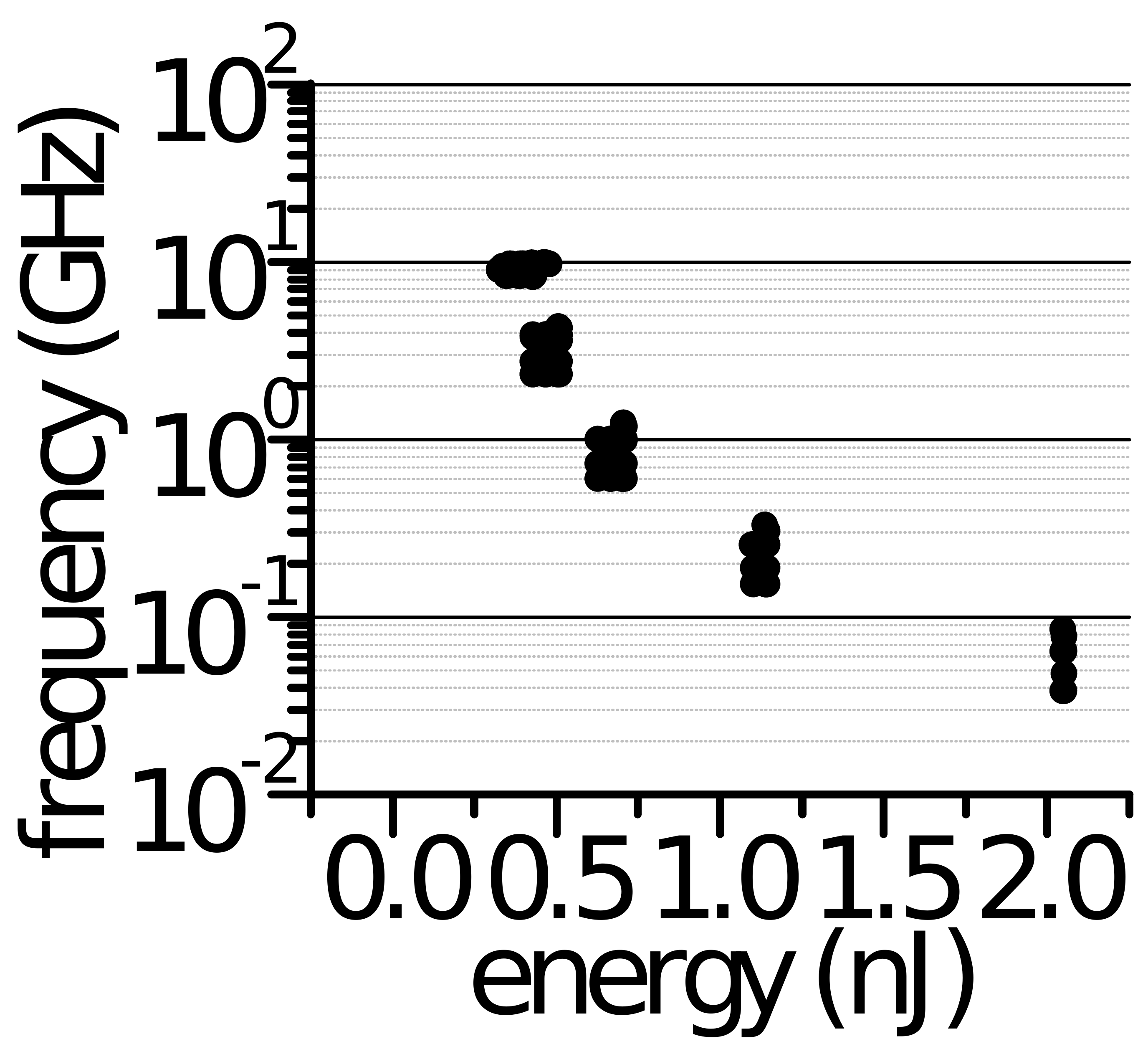}
   \label{f:sfq_latency_energy}
}
\hspace{-0.15in}
\subfigure[Area.]{
   \includegraphics[width=1.05in]{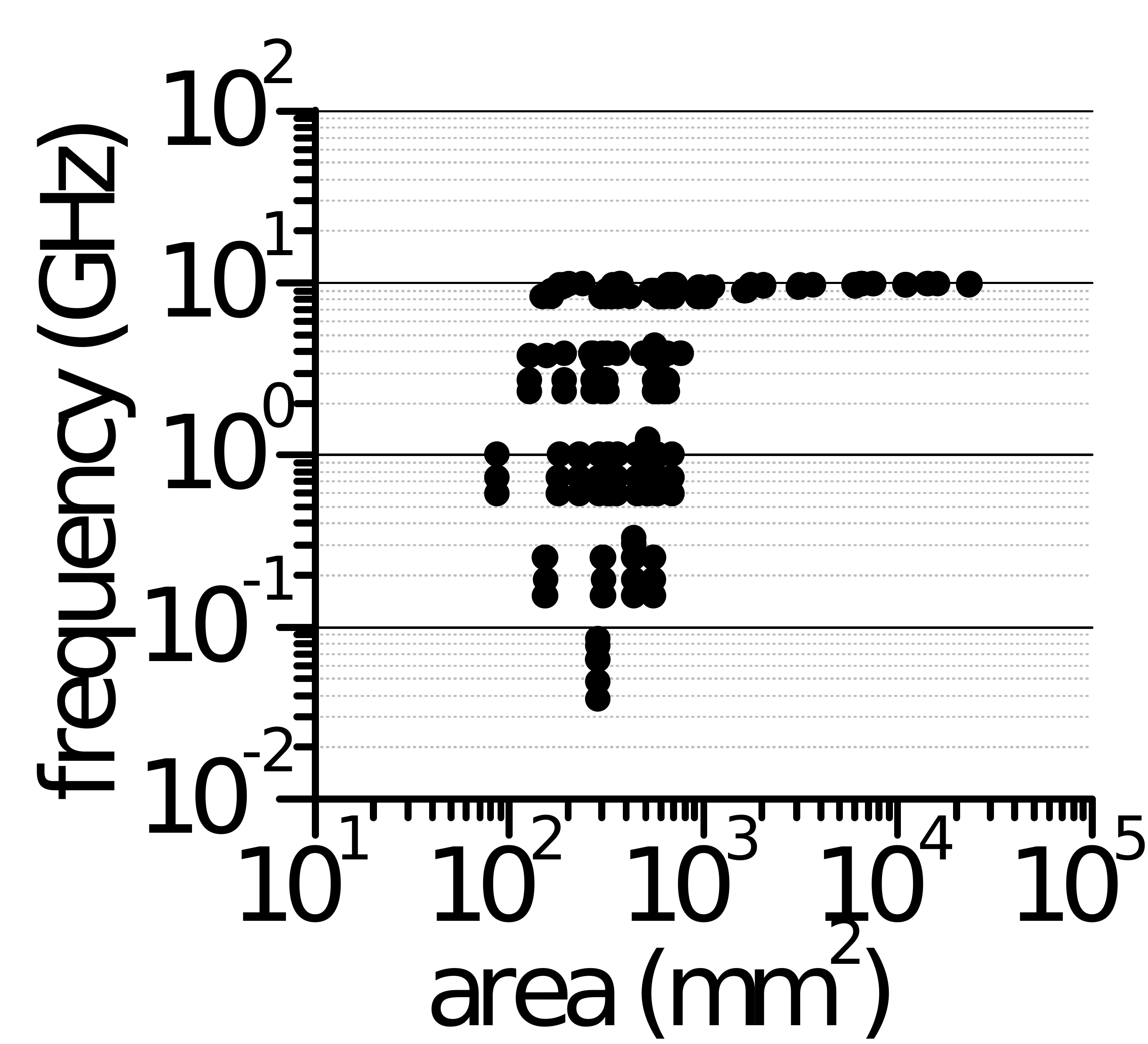}
   \label{f:sfq_latency_area}
}
\vspace{-0.15in}
\caption{The design space exploration.}
\label{f:sfq_design_exploration}
\vspace{-0.25in}
\end{figure}

\subsubsection{Pipeline Design Space Exploration}
\label{s:explore}

The design space exploration of our pipelined SFQ-CMOS array is exhibited in Figure~\ref{f:sfq_design_exploration}. The bottleneck of the entire pipeline of our SFQ-CMOS array lies in the stage of nTrons, whose latency is $\unit[103.02]{ps}$, since we cannot further break the latency into multiple pipeline stages. Therefore, the maximal frequency of our pipelined SFQ-CMOS array is $\unit[9.6]{GHz}$. To achieve the maximal pipeline frequency, we adjusted the size of CMOS sub-banks and the frequency of SFQ H-Trees. By reducing the size of CMOS sub-banks, the access latency to sub-banks is reduced to fit into one pipeline stage, since bitlines and wordlines in each MAT become shorter. However, the leakage power and area overhead of a sub-bank increased substantially, since more CMOS peripherals were added into each sub-bank. On the other hand, we inserted drivers and receivers to break a H-Tree into more pipeline stages, each of which has the latency of $\unit[103.02]{ps}$. As a result, both the area overhead and access energy of a pipelined SFQ-CMOS array increase.

\begin{table}[ht!]
\small
\setlength{\tabcolsep}{2pt}
\centering
\caption{The notations of the ILP formulation.}
\vspace{-0.1in}
\begin{tabular}{|c||l|}
\hline
Notation      & Description \\\hline\hline
\multirow{2}{*}{$\mathcal{M}$} & Memory object: weight ($\alpha$), input ($\beta$), output ($\gamma$), \\
                       & PSum ($\delta$) \\\hline
$i$                    & The $i_{th}$ edge in the DAG \\\hline
$ls$                   & SPM access: load ($\mathcal{L}$), and store ($\mathcal{S}$) \\\hline
\multirow{4}{*}{$st$}  & The status of $\mathcal{M}$: in a SHIFT array (\textit{H}), in a \\
                       & RANDOM array (\textit{R}), accesses between \textit{H} and \textit{R} \\
						           & (\textit{HR}), accesses between \textit{H} and DRAM (\textit{HD}), \\
							         & accesses between \textit{R} and DRAM (\textit{RD}) \\\hline
\end{tabular}
\label{t:sfq_variable_des}
\vspace{-0.1in}
\end{table}

\begin{figure}[ht!]
\centering
\includegraphics[width=3.3in]{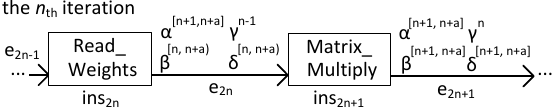}
\vspace{-0.15in}
\caption{The DAG of a convolutional layer.}
\label{f:sfq_data_graph}
\vspace{-0.15in}
\end{figure}

\subsection{A Compiler for Heterogeneous SPMs}

We built a novel compiler to allocate and prefetch memory objects onto SMART composed of SHIFT arrays and a RANDOM array for a SFQ systolic CNN accelerator by integer linear programming (ILP). No prior SPM management technique has the ability to schedule or prefetch memory requests for a systolic CNN accelerator, since prior work~\cite{Suhendra:RTSS2005,Udayakumaran:TECS2006,Deverge:ECRTS2007,Verma:TVLSI2006,Liu:ICCD2012} focuses on general-purpose applications with multiple basic blocks, each of which is an instruction sequence with no branches in except to the entry and no branches out except at the exit. A convolutional layer is a 6-nested loop~\cite{Chen:FPGA2015} belonging to a basic block. Our ILP-based compiler aims to allocate and prefetch memory objects at the instruction level without modifying the computing flow of a systolic CNN accelerator. Instead of 1-byte data, we set the granularity of allocation to memory objects, each of which is a multi-byte data block with consecutive addresses, to capture the temporal and spatial locality. Unlike prior SPM management schemes~\cite{Suhendra:RTSS2005,Udayakumaran:TECS2006,Deverge:ECRTS2007,Verma:TVLSI2006,Liu:ICCD2012}, which assume a memory object is alive throughout the whole basic block, we performed lifespan analysis of each memory object on the directed acyclic graph (DAG) of each convolutional layer to see how many iterations a memory object can live. Our compiler makes the near-optimal memory object allocation and prefetching to SMART on edges of the DAG of a convolutional layer. We designed our ILP-based technique for SMART consisting of private SHIFT arrays for inputs, weights, and PSums/outputs, and a shared RANDOM array for all, to enable data movements between SHIFT and RANDOM arrays, and to decide the schedule of a convolutional layer.

\textbf{Memory Object}: We considered weights ($\alpha$), inputs ($\beta$), outputs ($\gamma$), and PSum ($\delta$) results that need to be accumulated as candidates for SPM allocation. An ideal memory trace including all read and write accesses can be generated by the accelerator simulator SCALE-SIM~\cite{Samajdar:arXiv2018} by assuming that there is no delay caused by SPMs and DRAM. To capture fine-grained spatial and temporal locality, we grouped consecutive memory addresses across different processing elements (PEs) or consecutive cycles into one memory object $\mathcal{M}$. A memory object can be a weight filter kernel, a part of the input map, or an output channel.

\textbf{Lifespan Analysis}: We performed the lifespan analysis of memory objects at the instruction level on the DAG of a convolutional layer, as shown in Figure~\ref{f:sfq_data_graph}. Unlike prior SPM management schemes \cite{Suhendra:RTSS2005,Udayakumaran:TECS2006,Deverge:ECRTS2007,Verma:TVLSI2006,Liu:ICCD2012} compiling complex general-purpose applications on a CPU, our compiler focuses on each convolutional layer, which contains only one basic block. To maintain the original computing flow of the systolic CNN accelerator, a convolutional layer is first unrolled and compiled into a DAG. Each node in the DAG is an instruction of the systolic CNN accelerator, e.g., Google TPU~\cite{Jouppi:ISCA2017}, which has several types of CISC instructions as follows.

\begin{itemize}[nosep,leftmargin=*]
\item \textit{Read\_Weights}: Sending weights to the Matrix Unit.
\item \textit{Matrix\_Multiply}: Making the Matrix Unit perform a matrix multiply from the SPMs into accumulators.
\item \textit{Activate}: Performing activations and poolings.
\item \textit{Write(Read)\_Host\_Memory}: Writing (Reading) data from SPMs (the CPU memory) to the CPU memory (SPMs).
\end{itemize}
An edge between two instructions indicates that the destination node has data dependency on the source node. We annotated each edge with its related memory objects. For instance, at $e_{2n-1}$, i.e., the last edge of the $(n-1)_{th}$ iteration of the layer, the weight objects ($\alpha^{n}$) for the next ($n_{th}$) iteration have to be fetched.

\textbf{Prefetching}. Unlike prior SPM schemes~\cite{Suhendra:RTSS2005,Udayakumaran:TECS2006,Deverge:ECRTS2007,Verma:TVLSI2006,Liu:ICCD2012}, we enable the data fetching of memory objects that will be used in next several iterations by prolonging the lifespan of each memory object. For example, in Figure~\ref{f:sfq_data_graph}, for the first edge $e_{2n}$ of the $n_{th}$ iteration, besides writing the output objects of the previous $(n-1)_{th}$ iteration ($\gamma^{n-1}$), our compiler reads the weight objects $\alpha^{[n+1,n+a]}$ for next $a$ iterations, the input objects $\beta^{[n,n+a)}$ for current and next $(a-1)$ iterations, and the PSum objects $\delta^{[n,n+a)}$ for current and next $(a-1)$ iterations. The allocation and schedule results achieved by our ILP-based compiler are only ``near''-optimal, since we do not exhaustively search the best value of $a$. Instead, we set a fixed value for $a$.

\textbf{ILP Variable}: We define binary variables of the ILP formulas to attain the near-optimal scheme on a SFQ systolic CNN accelerator with SMART. As Table~\ref{t:sfq_variable_des} shows, these variables can be summarized as $\mathcal{M}^{i,st}_{ls}$, where $\mathcal{M}$ can be $\alpha$, $\beta$, $\gamma$, or $\delta$; \textit{ls} can be $\mathcal{L}$ or $\mathcal{S}$; and \textit{st} can be \textit{H}, \textit{R}, \textit{HR}, \textit{HD}, and \textit{RD}. For instance, if an input memory object is allocated to the SHIFT array on the $i_{th}$ edge of the DAG, we have $\beta^{i,H}=1$ and $\beta^{i,R}=0$. Setting a binary variable of SPM access to 1 indicates a load or store is enabled. For example, $\beta^{i,HD}_{\mathcal{L}}=1$ denotes loading the input memory object from the DRAM to the RANDOM SPM on the $i_{th}$ edge of the DAG.

\textbf{ILP Objective Function}: The objective function is to obtain the shortest execution time of each convolutional layer on a systolic CNN accelerator with heterogeneous SPM architecture. The objective function is summarized as
\begin{equation}
\begin{split}
& \max\sum_{i}\sum_{\mathcal{M}\in \{\alpha,\beta,\gamma,\delta\}} \{T_s^{H}\times \mathcal{M}^{i,H} + T_s^{R}\times \mathcal{M}^{i,R} \\
& - T_r^{HD}\times \mathcal{M}^{i,HD}_{\mathcal{L}} - T_r^{RD}\times \mathcal{M}^{i,RD}_{\mathcal{L}}  \\
& - T_r^{HR}\times \mathcal{M}^{i,HR}_{\mathcal{L}} - T_w^{HD}\times \mathcal{M}^{i,HD}_{\mathcal{S}} \\
& - T_w^{RD}\times \mathcal{M}^{i,RD}_{\mathcal{S}} - T_w^{HR}\times \mathcal{M}^{i,HR}_{\mathcal{S}}\}
\end{split}
\label{e:sfq_object_fun}
\end{equation}
where $T_s^{H}$ ($T_s^{R}$) is the reduced latency if a memory object is allocated to a SHIFT (RANDOM) array instead of the DRAM. $T_r^{HD}$ / $T_r^{RD}$ / $T_r^{HR}$ is the latency of reading a memory object from DRAM / DRAM / a RANDOM array and writing it to a SHIFT / RANDOM / SHIFT array. $T_w^{HD}$ / $T_w^{RD}$ / $T_w^{HR}$ is the latency of writing a memory object back to DRAM / DRAM / a RANDOM array from a SHIFT / RANDOM / SHIFT array.

\textbf{ILP Constraints}: We use the following ILP constraints to guarantee the correctness of the final SPM allocation and schedule of a convolutional layer.
\begin{itemize}[nosep,leftmargin=*]
\item \textbf{DAG and lifespan}: The scheduling and prefetching result has to match the lifespan analysis of memory objects, and the data dependency of the DAG.
\item \textbf{Consistency of SPM accesses}: The consistency of SPM accesses is enforced by
\begin{equation}
\begin{split}
& \forall i<j, \quad \mathcal{M}^{j,H} - \mathcal{M}^{j,HD}_{\mathcal{L}} - \mathcal{M}^{j,HR}_{\mathcal{L}} - \mathcal{M}^{i,H} = 0\\
& \forall i<j, \quad \mathcal{M}^{j,R} - \mathcal{M}^{j,RD}_{\mathcal{L}} - \mathcal{M}^{i,R} = 0\\
& \forall i<j, \quad \mathcal{M}^{j,HR}_{\mathcal{L}} - \mathcal{M}^{i,R} \leq 0
\end{split}
\label{e:sfq_object_spill}
\end{equation}
If we allocate a memory object to a SHIFT array on an edge $e_j$, as displayed in the first line of Equation~\ref{e:sfq_object_spill}, this memory object should be either allocated in the same array on a prior edge $e_i$ ($i<j$) or loaded to this SPM on edge $e_j$. The second line guarantees the consistency of SPM accesses in the RANDOM array. The last line enforces the memory object should be already allocated to the RANDOM array on edge $e_i$, if it is loaded to a SHIFT array on edge $e_j$ from this RANDOM array.

\item \textbf{SPM size}: The aggregate size of all memory objects allocated to the same array cannot exceed the array size.
\item \textbf{SPM bandwidth}: The total read (write) bandwidth of a SPM cannot exceed its maximal read (write) bandwidth.
\item \textbf{Sub-bank}: If two requests are scheduled to the same sub-bank at the same time, they are processed sequentially.
\end{itemize}

\begin{figure}[t!]
\centering
\begin{minipage}{.25\textwidth}
\centering
\includegraphics[width=1.7in]{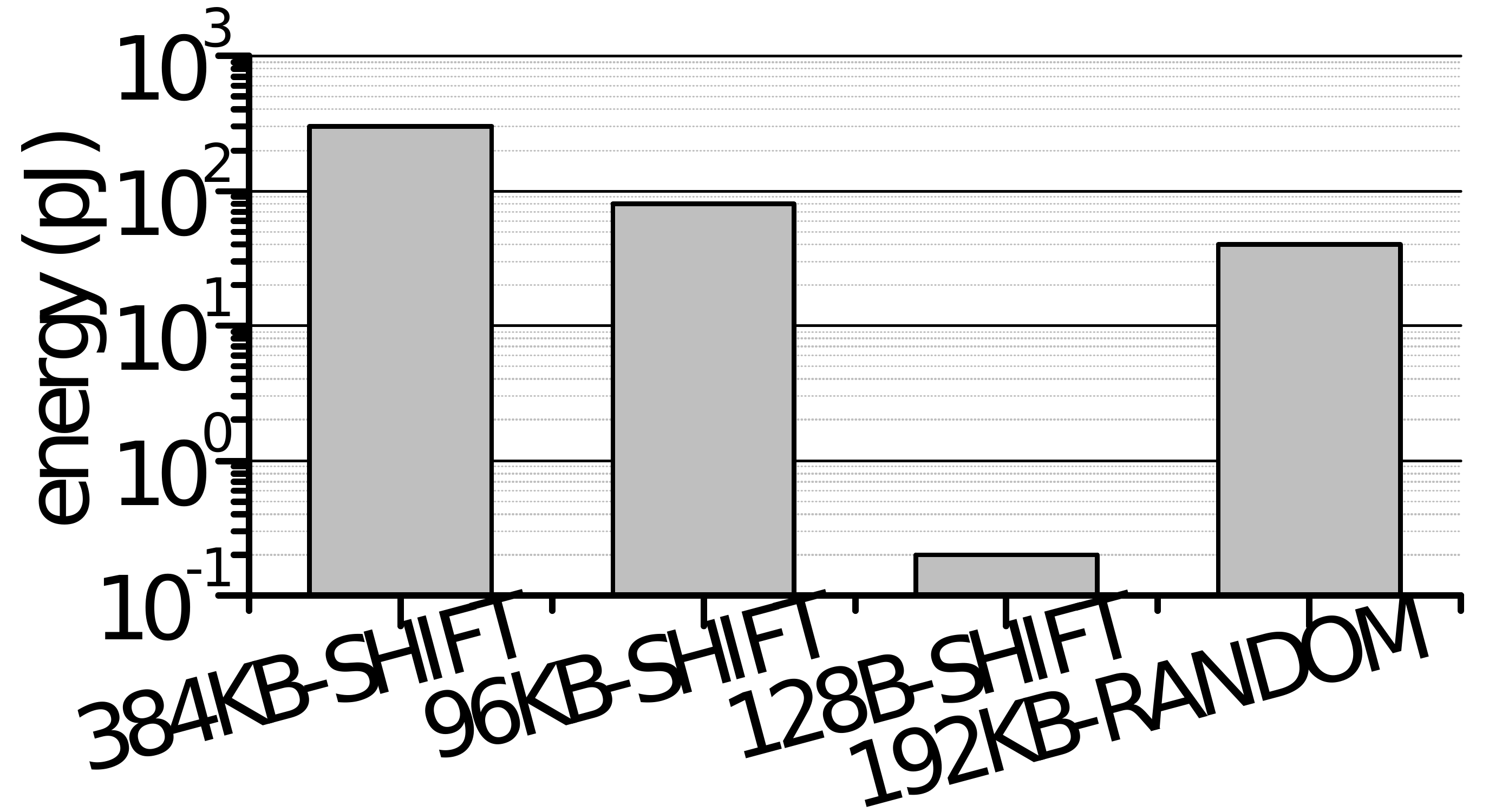}
\vspace{-0.1in}
\caption{The energy.}
\label{f:sfq_energy_overhead}
\end{minipage}%
\begin{minipage}{.25\textwidth}
\centering
\includegraphics[width=1.7in]{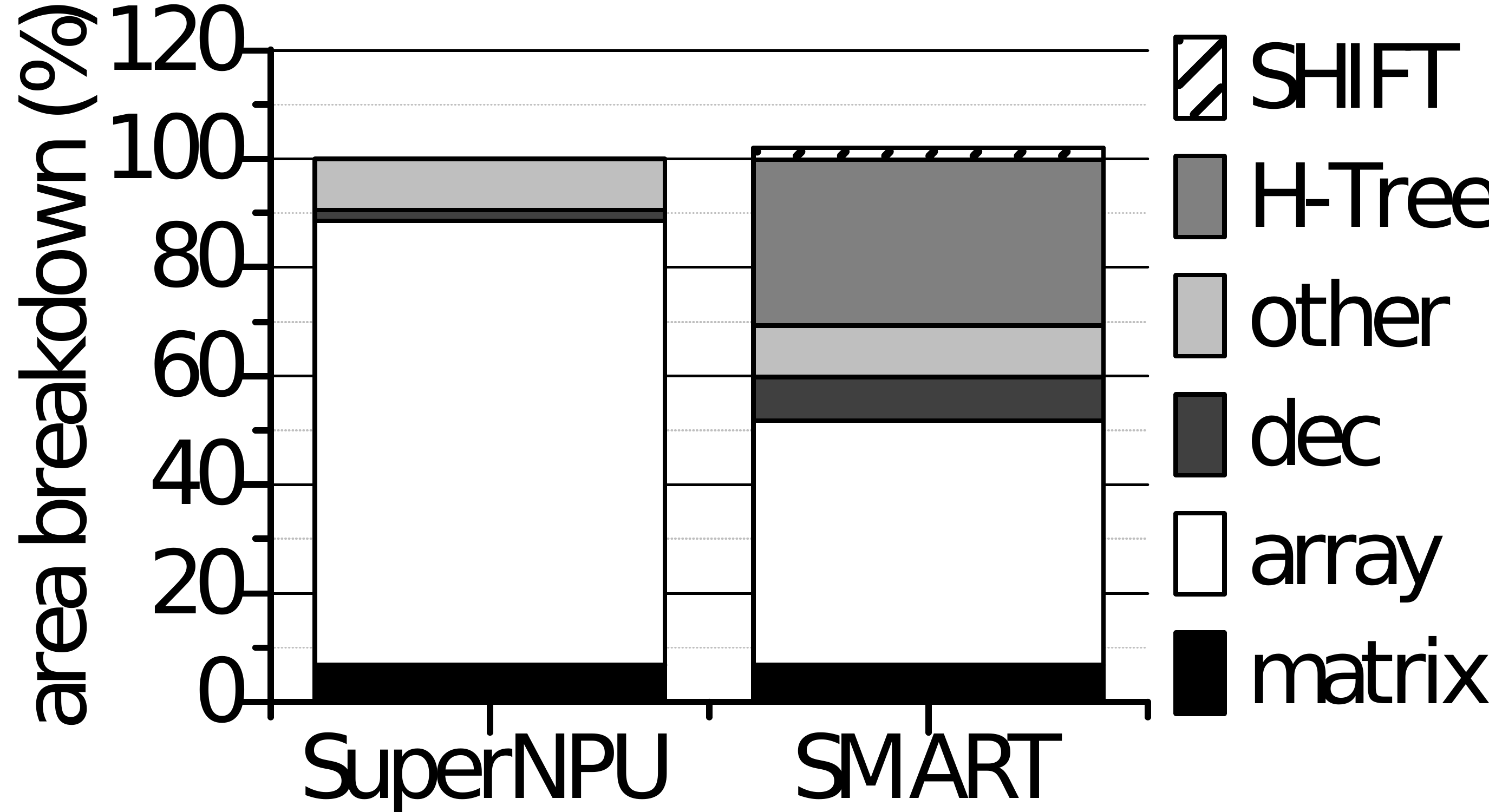}
\vspace{-0.1in}
\caption{The area.}
\label{f:sfq_area_overhead}
\end{minipage}%
\vspace{-0.2in}
\end{figure}

\subsection{Design Overhead}

\textbf{The Heterogeneous SPM}. SuperNPU~\cite{Ishida:MICRO2020} has a $\unit[24]{MB}$ 64-bank input SHIFT buffer, a $\unit[24]{MB}$ 256-bank output/PSum SHIFT buffer, and a $\unit[128]{KB}$ weight SHIFT buffer. In contrast, SMART has three 256-bank $\unit[32]{KB}$ SHIFT arrays for inputs, outputs/PSums, and weights, respectively. It also has a 256-bank $\unit[28]{MB}$ SFQ-CMOS SRAM array that can be operated at $\unit[9.7]{GHz}$ for all data. 
\begin{itemize}[nosep,leftmargin=*]
\item \textit{Latency}: The access latency of a SHIFT array is $\unit[0.02]{ns}$, while a SFQ-CMOS bank can read or write 1-byte data each $\unit[0.11]{ns}$.
\item \textit{Leakage}: A SHIFT array has no leakage, but the leakage power consumption of the pipelined SFQ-CMOS SRAM array is $\unit[102]{mW}$.
\item \textit{Dynamic energy}: As Figure~\ref{f:sfq_energy_overhead} shows, compared to a 384KB or 96KB bank of SuperNPU, the SHIFT arrays of SMART move only 128 DFFs per access, thereby reducing the access energy by 99\%. The access to the SFQ-CMOS array of SMART costs only 50\% of the dynamic energy of accessing the 96KB bank SuperNPU, due to low-power SFQ H-Trees.
\item \textit{Area}: Compared to SuperNPU, SMART reduces the SPM capacity by 41\%. But it has more CMOS sub-banks and more repeaters in SFQ H-Trees to achieve $\unit[9.7]{GHz}$. As Figure~\ref{f:sfq_area_overhead} shows, SMART increases the area by 3\%, when we assume SFQ JJs and CMOS transistors can be scaled to $28nm$~\cite{Ishida:MICRO2020}.
\end{itemize}

\textbf{The ILP-based Compiler}. We used SCALE-SIM~\cite{Samajdar:arXiv2018} to extract the DAGs of each CNN model, and identify memory objects. We adopted the Gurobi ILP solver~\cite{gurobi} to solve our ILP equations. For each of our CNN models (shown in Section~\ref{s:em}), the ILP solver can find a solution within one hour.

\begin{table}[t!]
\small
\setlength{\tabcolsep}{3pt}
\centering
\caption{The baseline configuration.}
\vspace{-0.1in}
\begin{tabular}{|l||l|}
\hline
Name                  & Description \\\hline\hline
\multirow{3}{*}{TPU}  & 0.7GHz; 45 TMAC/s peak perf.; PE array size \\
                      & $256\times256$; input, weight, and output: $\unit[24]{MB}$;\\
											& PSum: $\unit[4]{MB}$ \\\hline
                      & 52.6GHz; 842 TMAC/s peak perf.; PE array size\\
SuperNPU              & $64\times256$; input: 64-bank, $\unit[24]{MB}$; output/PSum: \\
                      & 256-bank, $\unit[24]{MB}$; weight: $\unit[128]{KB}$, $\unit[0.02]{ns}$\\\hline
\multirow{4}{*}{SMART}& 52.6GHz; 842 TMAC/s peak perf.; PE array size\\
                      & $64\times256$; three $\unit[32]{KB}$ SHIFT arrays for inputs, \\
                      & outputs/PSums, and weights: 256-bank, $\unit[0.02]{ns}$; \\
											& a $\unit[28]{MB}$ SFQ- CMOS array: 256-bank, $\unit[0.11]{ns}$\\\hline
\end{tabular}
\label{t:sfq_baseline_conf}
\vspace{-0.2in}
\end{table}

\begin{figure*}[t!]
\centering
\begin{minipage}{.5\textwidth}
\centering
\includegraphics[width=3.3in]{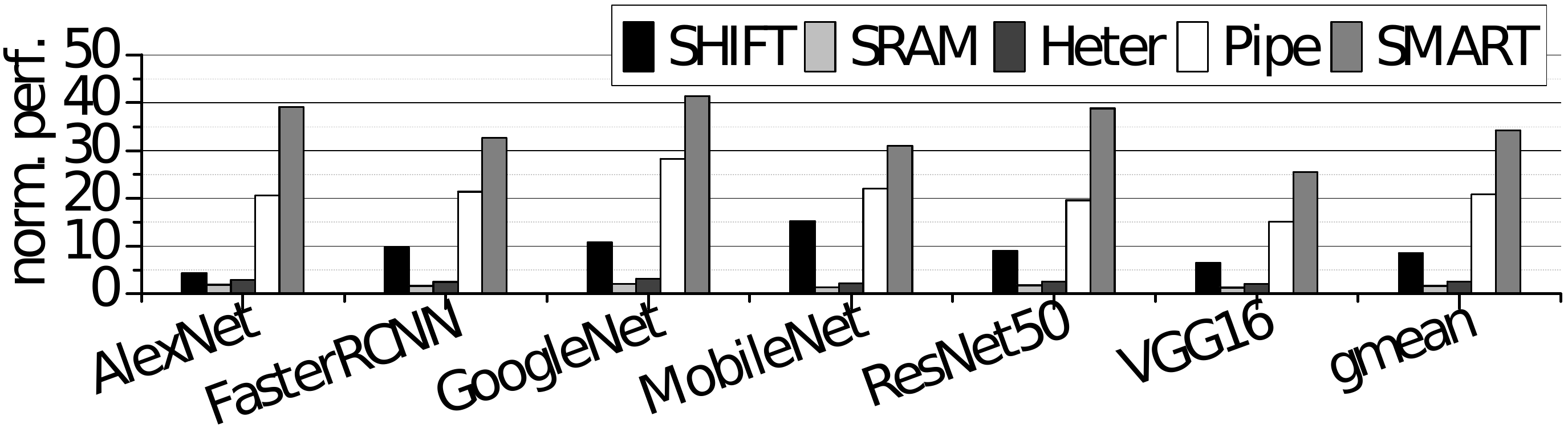}
\vspace{-0.15in}
\caption{The single-image speedup (norm. to TPU).}
\label{f:sfq_single_perf}
\end{minipage}%
\begin{minipage}{.5\textwidth}
\centering
\includegraphics[width=3.3in]{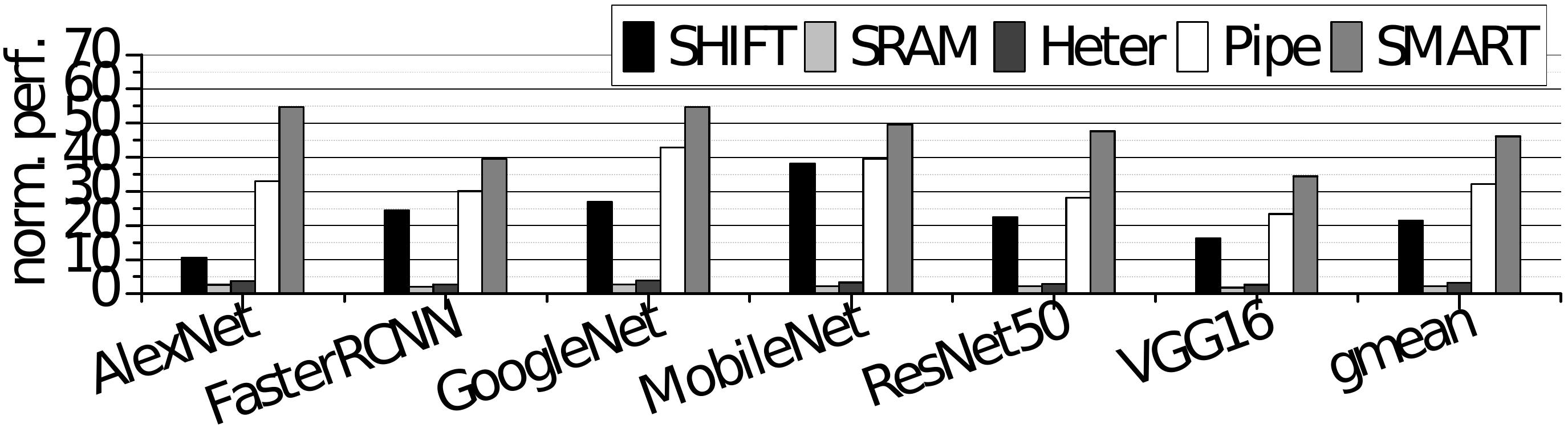}
\vspace{-0.15in}
\caption{The batch speedup (norm. to TPU).}
\label{f:sfq_batch_perf}
\end{minipage}%
\vspace{-0.1in}
\end{figure*}

\begin{figure*}[t!]
\centering
\begin{minipage}{.48\textwidth}
\centering
\includegraphics[width=3.3in]{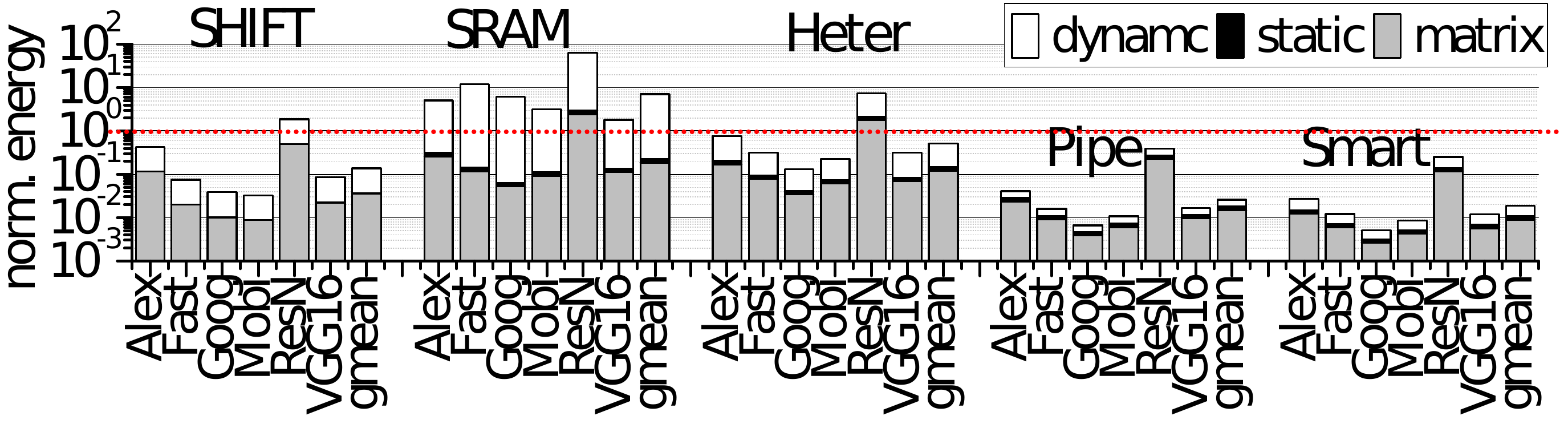}
\vspace{-0.15in}
\caption{The single image energy reduction (norm. to TPU; matrix: matrix unit energy; dynamic: SPM dynamic energy; and static: SPM static energy).}
\label{f:sfq_single_energy}
\end{minipage}%
\hspace{0.05in}
\begin{minipage}{.48\textwidth}
\centering
\includegraphics[width=3.3in]{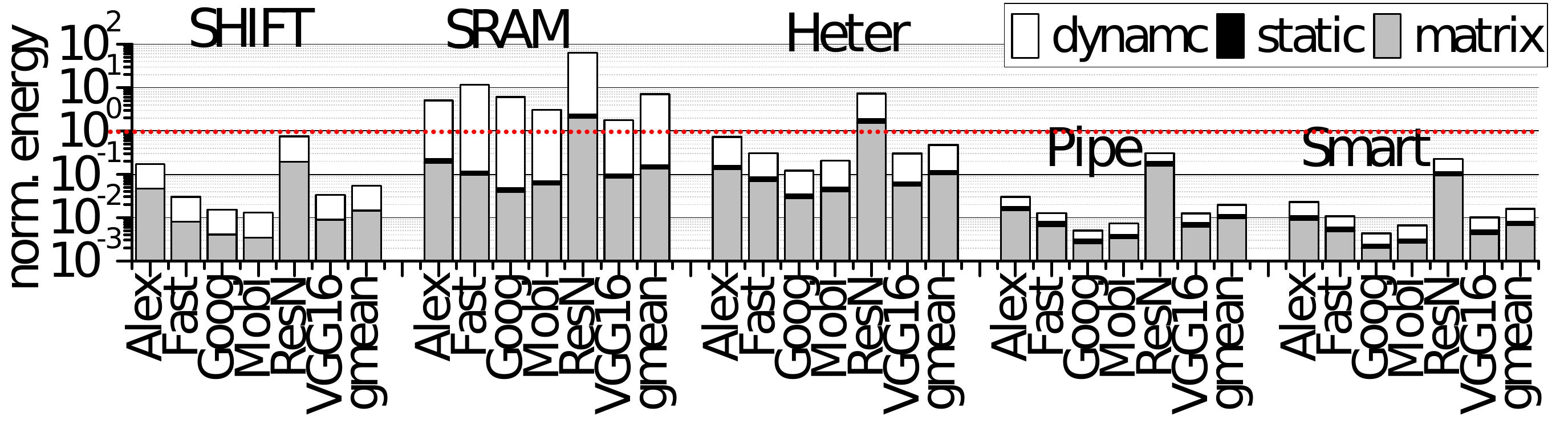}
\vspace{-0.15in}
\caption{The batch energy reduction (norm. to TPU; matrix: matrix unit energy; dynamic: SPM dynamic energy; and static: SPM static energy).}
\label{f:sfq_batch_energy}
\end{minipage}%
\vspace{-0.2in}
\end{figure*}

\section{Experimental Methodology}
\label{s:em}

\textbf{Simulation}. We used SCALE-SIM~\cite{Samajdar:arXiv2018} to model SMART, and our baselines including CMOS-based Google TPU~\cite{Jouppi:ISCA2017} and superconducting SFQ-based SuperNPU~\cite{Ishida:MICRO2020}. SCALE-SIM supports cycle-acc-urate performance simulations of a systolic CNN accelerator running inferences. The configurations of SMART and our baselines are shown in Table~\ref{t:sfq_baseline_conf}. We set the memory bandwidth of TPU, SuperNPU, and SMART to $\unit[300]{GB/s}$. The average power consumption of TPU is $40W$~\cite{Jouppi:ISCA2017}, while the power consumption of SuperNPU fabricated by the Hypres ERSFQ $1.0 \mu m$ technology~\cite{Yohannes:TAS2015} is only $1.9W$. We assume all components of SMART are also fabricated by the same ERSFQ $1.0 \mu m$ technology. The cooling cost of SuperNPU and SMART at 4K is $400\times$~\cite{Holmes:TAS2013} of their power consumption.

\textbf{CNN Models}. We selected six CNN models that have different characteristics, e.g., computational intensity, network topology and on-chip memory bandwidth needs. We ran single-image and batch-based inferences on baselines. The batch size setting is the same as~\cite{Ishida:MICRO2020}. For TPU and SMART, in a batch, AlexNet has 22 images, while VGG16 has 3 images. All the other models have 20 images in a batch. For SuperNPU, since it has larger SPMs, except VGG16 having 7 images in a batch, all the other models have 30 images in each batch. 

\textbf{Cryogenic Memory Modeling}. The details of SFQ-CMOS array modeling can be found in Section~\ref{s:pipeline}. We modified the cryogenic memory model cryo-mem~\cite{Lee:ISCA2019} to derive the access latency, energy consumption and area of VTM, MRAM, SNM arrays with the memory parameters in Table~\ref{t:sfq_mem_com}. We validated the simulated results of cryo-mem on VTM, MRAM, SNM arrays against their published array demonstrations~\cite{Semenov:TAS2019,Nguyen:SR2020,Butters:SST2021} respectively. We observed at most a 14\% error between the cryo-mem simulated data and the fabricated array. Compared to the large performance and energy degradation caused by VTM, MRAM, SNM arrays, the errors of cryo-mem are not significant.

\textbf{Schemes}. Besides our baseline TPU, we implemented and compared the following schemes:
\begin{itemize}[nosep,leftmargin=*]
\item \textit{SuperNPU}: The configuration of SuperNPU~\cite{Ishida:MICRO2020} is shown in Table~\ref{t:sfq_baseline_conf}.
\item \textit{SRAM}: SuperNPU replaces all SHIFT arrays by Josephson-CMOS SRAM arrays with the same capacity of TPU.
\item \textit{Heter}: Three $\unit[32]{KB}$ SHIFT arrays are added to the SRAM scheme. We assume an ideal SPM allocation, where the sequentially accessed data are always allocated in SHIFT arrays while the randomly accessed data are always allocated in the SRAM arrays.
\item \textit{Pipe}: Pipe replaces all Josephson-CMOS SRAM arrays of the Heter scheme by a $\unit[28]{MB}$ pipelined SFQ-CMOS SRAM array. 
\item \textit{SMART}: Our ILP-base compiler is used by the Pipe scheme. The prefetching iteration number $a$ is set to 3.
\end{itemize}

\begin{figure*}[bht!]
\centering
\begin{minipage}{0.23\textwidth}
\centering
\includegraphics[width=1.6in]{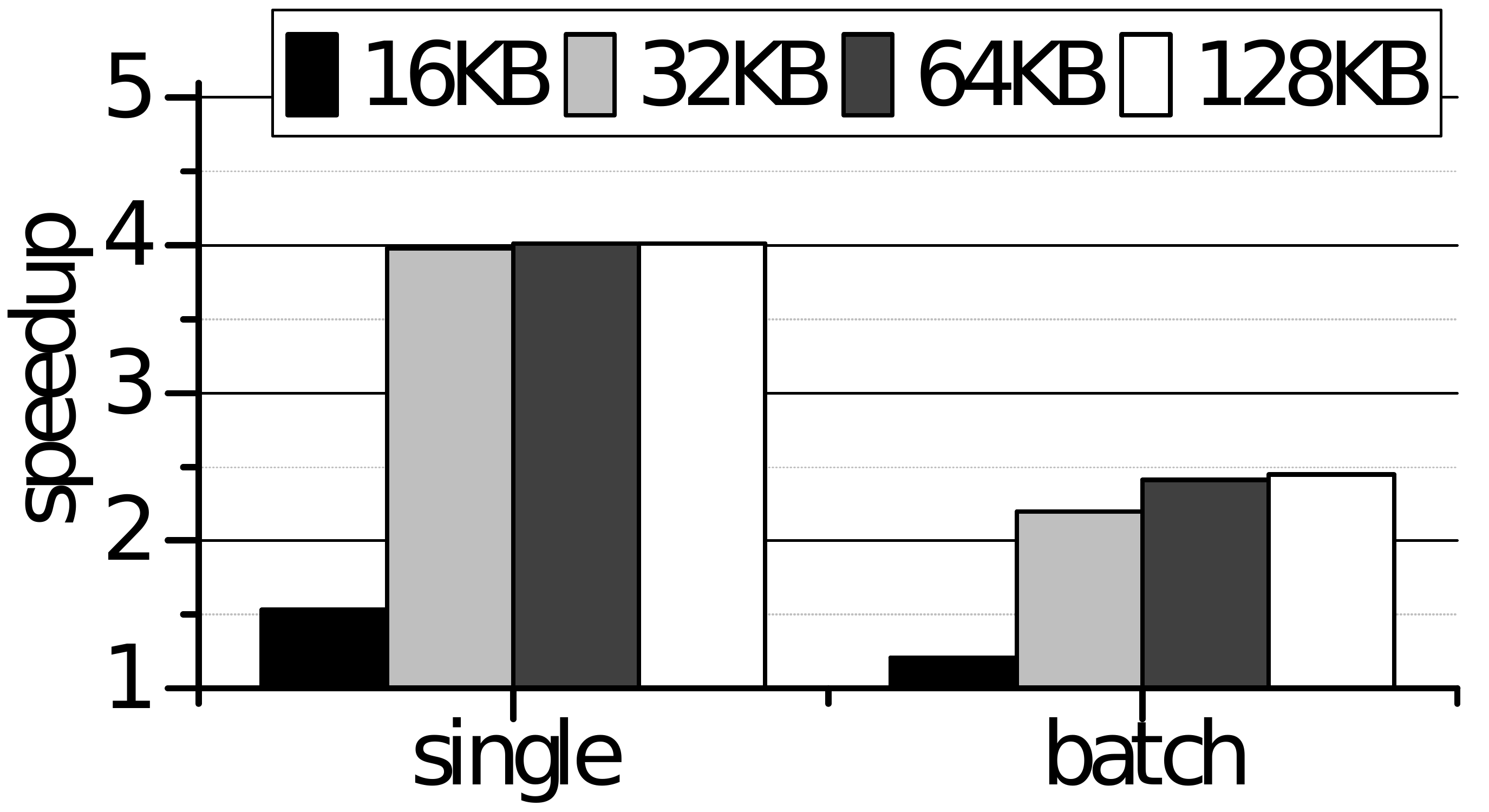}
\vspace{-0.3in}
\caption{SHIFT capacity (norm. to SuperNPU).}
\label{f:sfq_sen_shift}
\end{minipage}
\hspace{0.05in}
\begin{minipage}{0.23\textwidth}
\centering
\includegraphics[width=1.6in]{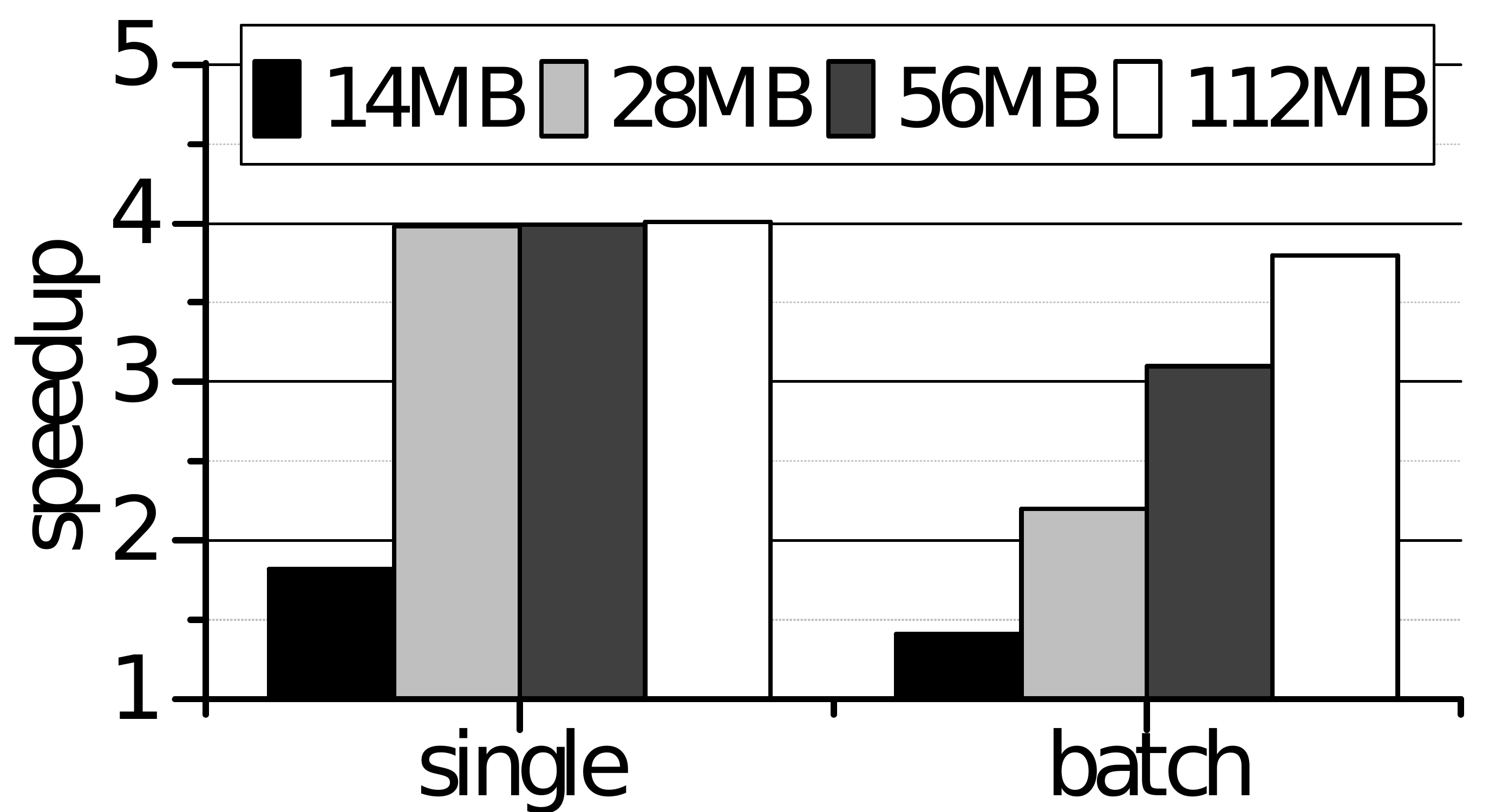}
\vspace{-0.3in}
\caption{RAND. capacity (norm. to SuperNPU).}
\label{f:sfq_sen_random}
\end{minipage}
\hspace{0.05in}
\begin{minipage}{0.23\textwidth}
\centering
\includegraphics[width=1.6in]{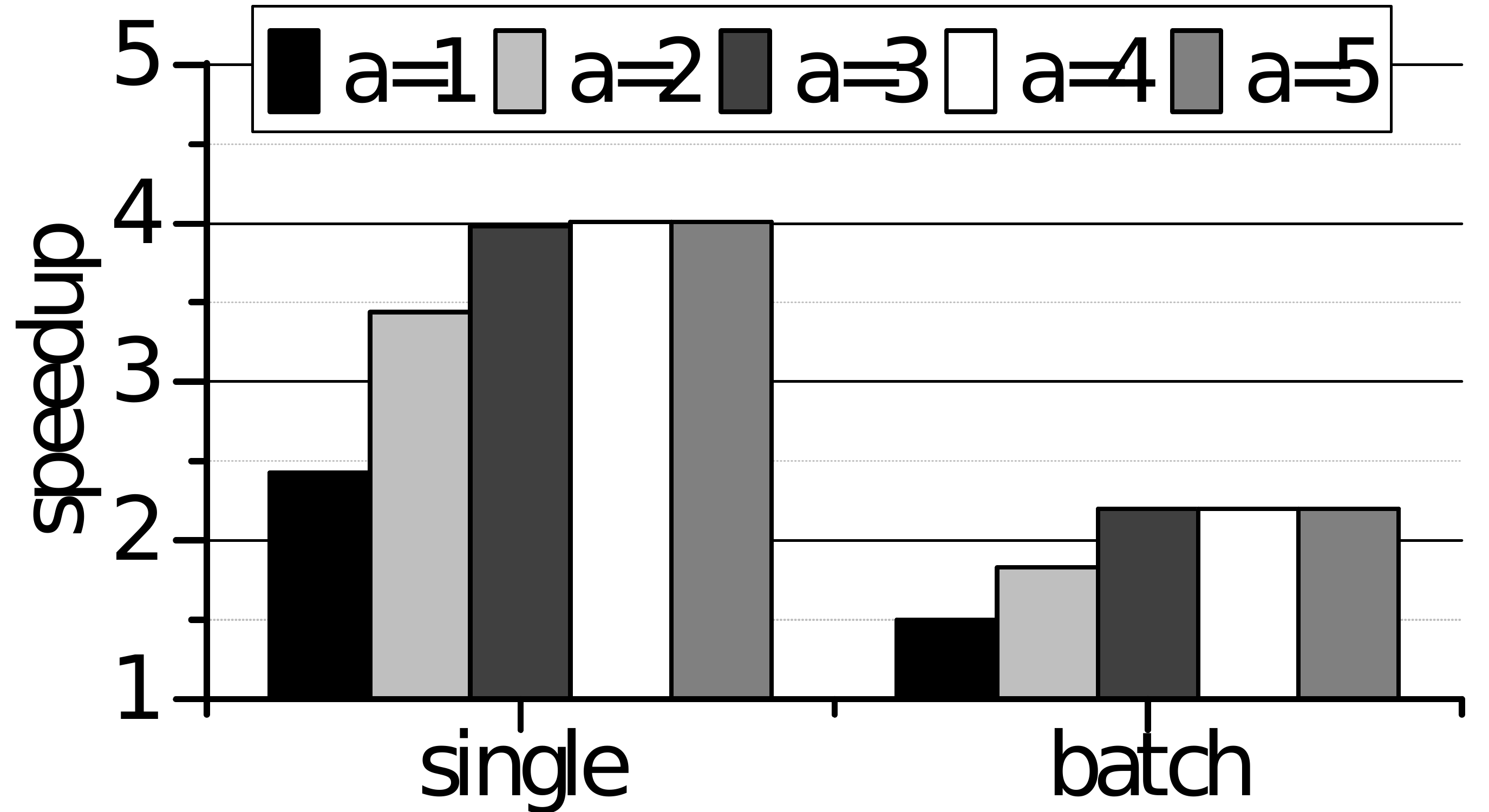}
\vspace{-0.3in}
\caption{Prefetch. iter. \# (norm. to SuperNPU).}
\label{f:sfq_sen_a}
\end{minipage}
\hspace{0.05in}
\begin{minipage}{0.23\textwidth}
\centering
\includegraphics[width=1.6in]{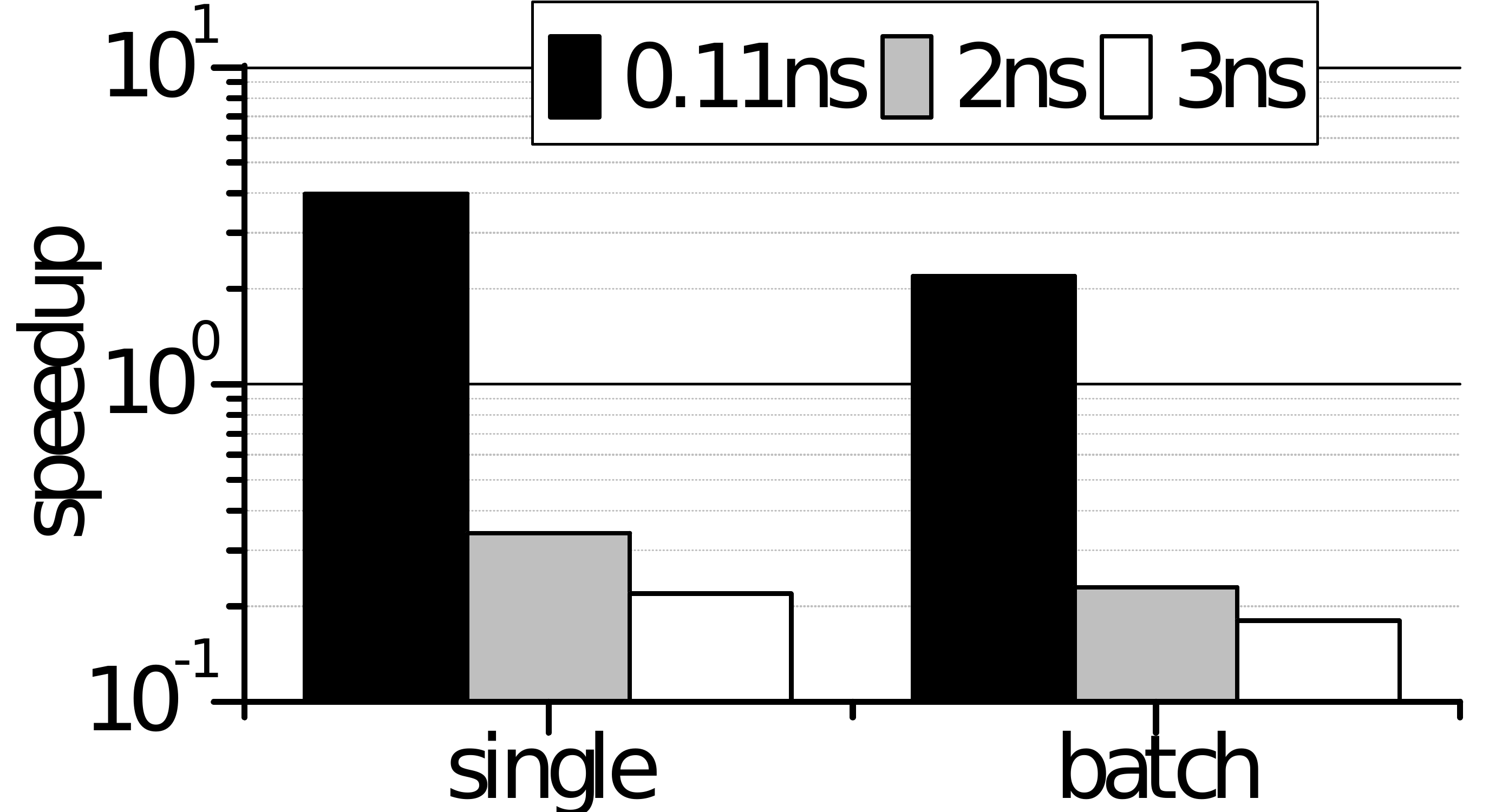}
\vspace{-0.3in}
\caption{RAND. W lat-ency (norm. to SuperNPU).}
\label{f:sfq_sen_write}
\end{minipage}
\vspace{-0.1in}
\end{figure*}

\vspace{-0.1in}
\section{Results and Analysis}
\label{s:result}

\subsection{Inferring a Single Image}

\textbf{Performance}. The performance improvement achieved by SMART inferring a single image is shown in Figure~\ref{f:sfq_single_perf}. The performance is measured by the throughput (i.e., TMAC/s) normalized to that of the TPU. Average customers are sensitive to the latency of their cloud-based machine learning services. Therefore, the performance of a single image inference becomes more critical, because TPUs in the cloud have no time to form a large image batch. For one-image inferences, SuperNPU improves the inference throughput by only $8.6\times$ over TPU, although the operating frequency of SuperNPU is $75\times$ higher than that of TPU. Compared to SuperNPU, Josephson-CMOS SRAM arrays actually decrease the inference throughput. This is because the benefit brought by the random access capability of Josephson-CMOS SRAM is offset by its slow access speed. Even if we add a small SHIFT array to each heterogeneous SPM, we cannot win back the performance loss. Heter still obtains lower inference throughput than SuperNPU. On the contrary, our pipe-lined SFQ-CMOS array (Pipe) greatly improves the inference throughput, on average, by $2.4\times$ over SuperNPU, due to its ultra-fast random access ability. Our ILP compiler (SMART) further increases the inference throughput improvement to $3.9\times$ over SuperNPU, since it enables the prefetching of input, weight, and PSum data of a model.

\textbf{Energy Consumption}. The energy comparison between various schemes when inferring a single image is shown in Figure~\ref{f:sfq_single_energy}. Since SuperNPU is fabricated by the ERSFQ technology, it has no leakage power. We consider the cooling cost of each scheme at 4K as $400\times$~\cite{Holmes:TAS2013} of the power consumption of that scheme. Since, on average, SuperNPU improves the performance per Watt by 23\% over TPU~\cite{Ishida:MICRO2020}, it consumes more energy on large CNN models, e.g., ResNet50 when considering the cooling overhead. SRAM and Heter tend to increase the inference energy when inferring a single image, because they obtain only longer inference latency and spend larger power in their Josephson-CMOS SRAM arrays. Our pipelined SFQ-CMOS array (Pipe) reduces the power consumption of RANDOM arrays by replacing CMOS H-Trees with SFQ H-Trees. Moreover, Pipe also shortens the inference latency over SuperNPU. As a result, Pipe reduces the inference energy by 81\%. SMART decreases the inference energy by 86\% over SuperNPU by further reducing the inference latency. On average, SMART uses only $1.9\%$ of the inference energy of TPU when inferring the same image. For SMART, 48\% of its energy is consumed by the matrix units, while 42\% of its energy is the dynamic energy of the heterogeneous SPM.

\subsection{Inferring a Batch of Images}

\textbf{Performance}. The performance improvement achieved by SMART inferring a batch of images is shown in Figure~\ref{f:sfq_batch_perf}. The inference performance of a batch of images shares the same trend as that of a single image. Compared to the single image case, SuperNPU inferring a batch of images improves the inference throughput by $2.5\times$. In contrast, SMART processing a batch of images improves the inference throughput by only $34.5\%$ over the single image case of SMART. This is because SuperNPU has larger on-chip space to store more images, i.e., SuperNPU has $\unit[48]{MB}$ SPM arrays, while SMART has only a $\unit[28]{MB}$ on-chip RANDOM array. On average, when processing a batch of images, SMART improves the inference throughput over SuperNPU by $2.2\times$.

\textbf{Energy Consumption}. The energy reduction of SMART inferring a batch of images is shown in Figure~\ref{f:sfq_batch_energy}. We also consider the cooling cost in the comparison. The inference energy of a batch shares the same trend as that of a single image. On average, SMART reduces the inference energy by 71\% over SuperNPU, and uses only 1.6\% of the inference energy of TPU when processing a batch of images. In SMART, 42.3\% of its energy consumption is the energy of the matrix units, while 48.9\% of the energy is the dynamic energy of its heterogeneous SPM arrays.

\subsection{Sensitivity Study}

\textbf{SHIFT array capacity}. The sensitivity study on the capacity of SHIFT arrays in SMART is shown in Figure~\ref{f:sfq_sen_shift}. The input, output/PSum, and weight data have three SHIFT arrays with the capacity of $X$, where $X$ can be $\unit[16]{KB}$, $\unit[32]{KB}$, $\unit[64]{KB}$, and $\unit[128]{KB}$. Compared to $\unit[32]{KB}$, the larger capacity of SHIFT arrays cannot help single-image inferences, and only slightly improve the inference throughput on a batch of images by 11\%. On the contrary, three $\unit[16]{KB}$ SHIFT arrays greatly increase the swapping traffic between SHIFT arrays and the RANDOM array, thereby decreasing the inference throughput of a single image and a batch of images by 61\% and 45\%, respectively.

\textbf{RANDOM array capacity}. The sensitivity study on the RANDOM array capacity in SMART is shown in Figure~\ref{f:sfq_sen_random}. Though the input, output/PSum, and weight data have three SHIFT arrays respectively, they share the same RANDOM array. We tried different capacities of the RANDOM array in the figure. Compared to $\unit[28]{MB}$, further increasing the RANDOM array capacity does not improve the single-image inference throughput. However, a $\unit[56]{MB}$ ($\unit[112]{MB}$) RANDOM array improves the inference throughput of a batch by 41\% (73\%). On the other hand, a smaller RANDOM array hurts the inference throughput of both a single image and a batch of images.

\textbf{Prefetching iteration number}. The sensitivity study on the prefetching iteration number of SMART is shown in Figure~\ref{f:sfq_sen_a}. Our ILP compiler achieves only near-optimal results, since we did not exhaustively explore the optimal prefetching iteration number. We set the prefetching iteration number $a=3$. $a=1$ indicates there is no prefetching. A smaller $a$ substantially decreases the throughput of both single-image and batch inferences. On the other hand, a larger $a$ (e.g., $a=4$) does not obviously improve the inference throughput of six CNN models we selected.

\textbf{Write latency}. The sensitivity study on the write latency of the RANDOM array in SMART is shown in Figure~\ref{f:sfq_sen_write}. Since MRAM and SNM have smaller cell sizes than SRAM, if JJs can be scaled to the same size of a transistor, it is possible to use them to build a much denser RANDOM array. However, their write latency is much longer. We explore different values of the write latency of the RANDOM array in the figure. A longer write operation significantly decreases the throughput of both single-image and batch inferences, since the outputs of a layer are the inputs of the next layer. Therefore, these high-density cryogenic memory technologies may not be ideal candidates to implement the RANDOM array due to their slow writes.

\section{Related Work}
\label{s:related}

\textbf{SFQ Accelerators}. As we are approaching the end of Moore's Law, several ambitious designs for superconducting ALUs~\cite{Kirichenko:TAS2019,Filippov:TAS2011} and microprocessors~\cite{Yoshikawa:TAS2003} have been presented to demonstrate the capability of SFQ computing. For domain-specific computing, besides SFQ CNN systolic accelerators, a SFQ stochastic-computing-based deep learning accelerator~\cite{Cai:ISCA2019} also demonstrates ultra-high inference throughput. Moreover, a SFQ-based temporal logic accelerator~\cite{Tzimpragos:ASPLOS2020} is built to significantly boost the throughput of genome alignment. A SFQ-based SHA-256 accelerator~\cite{Tannu:CF2019} is designed to maximize the processing throughput of cryptographic hash functions. These superconducting designs primarily depend on simplified architectures, bit-serial processing, and shift registers. However, the use of SFQ shift registers is not a viable solution for more complex accelerator designs.

\textbf{Cryogenic Memories and Caches}. Recent work adopts the 77K cryogenic temperature to improve the performance and energy consumption of off-chip DRAM main memories~\cite{Lee:ISCA2019} and on-chip SRAM caches~\cite{Min:ASPLOS2020}. However, these studies investigate only how the main memory and cache architectures are influenced by the 77K temperature when running general-purpose applications on CPUs. No prior work designs an on-chip SPM architecture for SFQ systolic CNN accelerators at the 4K temperature.

\section{Conclusion}
\label{s:con}

In this paper, we propose a heterogeneous SPM architecture, SMART, consisting of SHIFT arrays and a RANDOM array for SFQ deep learning accelerators to maximize their inference throughput. However, we found that no existing memory technology can serve as the RANDOM array of SMART to obtain high inference throughput, small chip area, and low power consumption at the same time. We built a fast, dense and power-efficient pipelined CMOS-SFQ array that supports random accesses in SMART. We also created an ILP-based SPM allocation and prefetching technique to minimize the inference latency on SMART. Experimental results show that, with the same area overhead, compared to the prior SHIFT-based SFQ CNN accelerator, SMART improves the inference throughput by $3.9\times$ ($2.2\times$), and reduces the inference energy by $86\%$ ($71\%$) when inferring a single image (a batch of images).

\begin{acks}
The authors would like to thank the anonymous reviewers for their valuable comments and helpful suggestions. This work was partially supported by the National Science Foundation (NSF) through awards CCF-1908992, CCF-1909509, and CCF-2105972.
\end{acks}

\bibliographystyle{ACM-Reference-Format}
\bibliography{super}

\end{document}